\documentclass[a4paper,11pt]{article}

\usepackage[a4paper,left=2.73cm,right=2.7cm,top=3cm,bottom=3.5cm]{geometry}
\usepackage{amsmath,amssymb,graphicx,colortbl,xcolor,caption,subcaption,wrapfig}
\usepackage[colorlinks=true,linktocpage=true,linkcolor=blue,citecolor=blue]{hyperref}

\usepackage[all]{xy}
\addtocontents{toc}{\protect\setcounter{tocdepth}{1}}

\def\d{\mathrm{d}}
\newcommand*\rfrac[2]{{ #1}  / { #2}}

\numberwithin{equation}{section}

\newcommand{\ud}{U_\mt{dual}}
\newcommand{\udp}{U_\mt{dual}'}
\newcommand{\up}{U_\mt{pert}}

\newcommand{\eqq}[1]{(\ref{#1})}
\newcommand{\leff}{g_\mt{eff}}
\newcommand{\lp}{\ell_\mt{P}}
\newcommand{\uh}{u_\mt{H}}
\newcommand{\rh}{r_\mt{H}}
\newcommand{\ul}{U_\mt{lower}}
\newcommand{\uu}{U_\mt{upper}}
\newcommand{\uc}{U_\mt{cross}}
\newcommand{\ls}{\ell_s}
\newcommand{\nq}{n_q}
\newcommand{\mt}[1]{\textrm{\tiny #1}}
\newcommand{\nc}{N}

\newcommand{\gym}{g_\mt{YM}}

\newcommand{\sac}{\, , \qquad}
\newcommand{\eq}[1]{(\ref{#1})}

\newcommand{\be}{\begin{equation}}
\newcommand{\ee}{\end{equation}}
\newcommand{\bal}{\begin{align}}
\newcommand{\bse}{\begin{subequations}}
\newcommand{\ese}{\end{subequations}}
\newcommand{\bea}{\begin{eqnarray}}
\newcommand{\eea}{\end{eqnarray}}

\begin{document}

\begin{titlepage}

\hfill{ICCUB-14-061}

\vspace{1cm}
\begin{center}

{\LARGE{\bf (Super)Yang-Mills at Finite Heavy-Quark Density}}

\vskip 45pt
{\large \bf Ant\'on F. Faedo$^{1}$, Arnab Kundu$^{1}$, David Mateos$^{1,2}$ and Javier Tarr\'\i o$^{1}$}

\vskip 10pt
{$^{1}$Departament de F\'\i sica Fonamental and Institut de Ci\`encies del Cosmos, \\ Universitat de Barcelona, Mart\'\i\  i Franqu\`es 1, ES-08028, Barcelona, Spain.}\\

\vskip 10pt
{$^{2}$Instituci\'o Catalana de Recerca i Estudis Avan\c cats (ICREA), \\
Passeig Llu\'\i s Companys 23, ES-08010, Barcelona, Spain}

\vskip 10pt
\end{center}

\vspace{10pt}
\abstract{\normalsize
We study the  gravitational  duals of $d$-dimensional Yang-Mills theories with $d\leq 6$ in the presence of an ${\cal O} (\nc^2)$ density of heavy quarks, with $\nc$ the number of colors. For concreteness we  focus on maximally supersymmetric Yang-Mills, but our results apply to a larger class of theories with or without supersymmetry. The gravitational solutions  describe renormalization group flows towards infrared scaling geometries characterized by fixed dynamical and hyperscaling-violating exponents. The special case $d=5$ yields an $AdS_3 \times \mathbb{R}^4 \times S^4$ geometry upon uplifting to M-theory. We discuss the multitude of physical scales that separate different dynamical regimes along the flows, as well as the validity of the supergravity description. We also present exact black brane solutions that encode the low-temperature thermodynamics.
}

\end{titlepage}

\tableofcontents

\hrulefill
\vspace{15pt}

\section{Introduction}
The study of Quantum Field Theory in different dimensions is compelling for several reasons. For particle physics it is a useful exercise that can shed light, provide intuition, and place into a more general context, the physically  relevant case $d=4$. For condensed matter applications, the cases $d=2,3$ are of direct phenomenological interest. In the case of strongly coupled systems with a finite density of quarks or electrons, the number of non-perturbative tools is very limited. In this context, the gauge/string duality \cite{Maldacena:1997re}  provides a stimulating set of toy models in which a first-principle description is possible. 

With this motivation in mind, in this paper we will study the finite-density physics of a class of quantum field theories with known gravitational duals. As we will explain in the Discussion section, our results apply supersymmetric and non-supersymmetric theories. However, for concreteness, in most of the paper we will focus on the simple cases of\footnote{In this paper $d$ and $p$ will always denote the number of spacetime and space dimensions in the gauge theory, respectively.} \mbox{$d=p+1$} dimensional, $SU(\nc)$, maximally supersymmetric super Yang-Mills (SYM) gauge theories that are realized on the worldvolume of $\nc$ coincident D$p$-branes in type II string theory \cite{Itzhaki:1998dd}.\footnote{Throughout this paper we will always assume that $\nc$ is large.} Unlike quarks or electrons, the matter in these theories is not in the fundamental representation of the gauge group. However, fundamental matter can be added by introducing  an additional set of so-called flavor branes \cite{Karch:2002sh}. The resulting system is a theory with dynamical matter in the fundamental representation, to which we will loosely refer as `quarks' henceforth. Turning on a  quark density corresponds on the gravity side to adding fundamental strings dissolved inside the flavor branes \cite{Kobayashi:2006sb}. The supergravity description of these systems  is complicated by the fact that, generically,  one must take into consideration the backreaction of both the flavor branes and the strings (see e.g.~\cite{Bigazzi:2011it}). For this reason, in this paper we will take a simplifying limit in which the mass of the quarks is sent to infinity. In this limit the flavor branes disappear and one is left with the backreaction of fundamental strings stretching all the way to the boundary of the spacetime \cite{Rey:1998ik,Maldacena:1998im}. On the gauge theory side the heavy quarks become non-dynamical and the quark number cannot fluctuate. We emphasize that we are interested in quark densities of ${\cal O}(\nc^2)$ since we would like the backreaction of the strings to be captured by  classical gravity. We will show that the resulting dual solutions describe renormalization group (RG) flows towards infrared (IR) geometries characterized by fixed dynamical and hyperscaling-violating exponents. Despite our infinite-mass limit, preliminary investigations actually indicate that these geometries also play a role in the case of finite-mass, dynamical quarks \cite{us}. 

For each string an orientation along the internal, compact  part of the geometry must be chosen. In the dual gauge theory this direction specifies the quantum numbers of the quark under the global symmetries of the theory (some of which may be R-symmetries in supersymmetric cases). We will chose to distribute the strings homogeneously along the compact directions of the ten-dimensional geometry, and we will refer to this procedure as `smearing'. Roughly speaking, in the gauge theory this means that we introduce quarks with all possible quantum numbers. This allows us to reduce the supergravity equations to ordinary (as opposed to partial) differential equations, while presumably leaving the qualitative IR physics unchanged with respect to other possible choices. This is suggested e.g.~by the results of Refs.~\cite{Kumar:2012ui} and \cite{Faedo:2013aoa}, which found that different string smearings gave rise to the same IR geometries. In some supersymmetric theories (e.g.~in the SYM theories that we will focus on) a single quark, or a group of quarks oriented all in the same direction, will generically preserve a fraction of the supersymmetry. In contrast, the smeared configuration will typically break it completely.   

In the case $d=4$, the strategy that we have just described was implemented in \cite{Kumar:2012ui}. The dual gauge theory is the conformally-invariant ${\cal N}=4$ SYM theory. Ref.~\cite{Kumar:2012ui} found that the backreaction of the strings induces an RG flow between $AdS_5\times S^5$ in the ultraviolet (UV) and an IR geometry\footnote{This geometry appeared previously in \cite{Azeyanagi:2009pr}.}
 invariant under the Lifshitz scaling 
\be
\label{eq.Lifshitzscaling}
t \to \Lambda^z \, t \ , \qquad x^i \to \Lambda\, x^i \,,
\ee
with dynamical exponent $z=7$. Although this exact invariance of the IR metric is mildly broken by a logarithmically-running dilaton, we will still refer to these type of solutions as `fixed points'. 

If $d\neq 4$  the SYM theory is not conformally invariant but is characterized by a dimensionful coupling constant $\gym^2$ with dimensions of (length)$^{d-4}$. The main result of our paper is that in all non-conformal cases with\footnote{As is well known, the cases with $d>6$ are problematic for several reasons and will not be considered here. In addition, in these cases the strings' backreaction modifies the UV asymptotics of the corresponding D$p$-brane solution. The case $d=1$ corresponds to D0-branes and must be treated separately, as we explain in Section \ref{sec.D0branes}.} $2\leq d\leq 6$ the presence of the strings induces an RG flow towards a Hyperscaling-violating Lifshitz (HV-Lif) geometry whose non-compact part takes the form
 \be 
 \label{eq.genericHVLifshitz}
\d s^2 =  \left( \frac{r}{L} \right)^{-\frac{2\,\theta}{p}} \left[ - \left( \frac{r}{L}\right)^{2\, z}  \d t^2 + \left( \frac{r}{L}\right)^{2} \d x_p^2 + \left( \frac{L}{r}\right)^{2} \d r^2 \right] \,,
\ee
with the dynamical and the hyperscaling-violating exponents given respectively by 
\be
z = \frac{16-3\, p}{4-p} \ , \qquad \theta = \frac{p\, (3-p)}{4-p}  \ .
\ee
In the full solution, the metric above is accompanied by  logarithmically-running scalars. The case $p=4$ can be understood as a well defined limit of the $p\neq 4$ formulas and results in a geometry  whose non-compact part is conformal to $AdS_2\times \mathbb{R}^4$ in Type IIA supergravity, and which becomes exactly $AdS_3 \times \mathbb{R}^4$ upon uplifting to M-theory.

Under the scaling transformation \eqref{eq.Lifshitzscaling}, together with $r \to \Lambda^{-1} \, r $, the  line element \eq{eq.genericHVLifshitz} transforms as
\be
ds \to \Lambda^{\theta/p} \, ds \ .
\label{as}
\ee
As a consequence, thermodynamic quantities such as the free energy or the entropy density are expected to scale at low temperatures as 
\be
\label{thermo}
F\sim T^\frac{p+z-\theta}{z} \sac s \sim T^\frac{p-\theta}{z} \,, 
\ee
respectively. We will verify this explicitly by constructing the `blackened' supergravity solutions dual to the gauge theory at finite temperature.  

Metrics of the form \eqref{eq.genericHVLifshitz} are known to present several generic issues that of course also afflict our solutions. Near the origin ($r\to 0$) there are divergent tidal forces, and for a non-vanishing HV exponent the curvature invariants diverge either close to or far from the origin. On the other hand, all our solutions verify the two inequalities 
\be\label{eq.NECs}
(p-\theta) \Big[ p(z-1)-\theta \Big]\geq0 \ , \qquad (z-1)(p+z-\theta)\geq0 \,,
\ee
that guarantee that the null energy condition (NEC) is not violated, as expected from the fact that the stress-energy tensor sourcing our solutions is a sensible one, namely that of fundamental strings.

The crossover between the UV solution, determined by the  D$p$-branes of type II string theory, and the HV-Lif solution in the IR, where the backreaction of the strings is important, occurs at a dimension-dependent energy scale.
The comparison between this scale and the ones present in the uncharged solution allows us to classify different types of RG flows:  for $d<4$ there are four  classes, for $d>4$ there are three and for $d=4$ only one, due to the conformality of the uncharged field theory.
Not all of these are captured by our solutions. For $d<4$ ($d>4$) extremely large (small) values of the quark density lead to RG flows that should be described by perturbative YM theory, therefore a supergravity description is not needed.
Conversely, for very small (large) values of the quark density the validity of the type II solution breaks down due to the dilaton becoming large. In these cases one needs a dual description (an S-dual one in Type IIB and an uplift to M-theory in Type IIA) to correctly describe the strongly coupled system. We will discuss some of these dual configurations.

This paper is organized as follows.
In section \ref{review} we review the gravity dual of SYM theories for any number of dimensions. 
In particular we describe the several relevant scales  and present our conventions for the supergravity actions.

In section \ref{sec.solutions} we discuss how the presence of string sources determines our setup and the action of the system.
Then we discuss how the D$p$-brane solution is modified asymptotically in the UV by the presence of strings, and exhibit a new fixed point solution given by a metric of the form \eqref{eq.genericHVLifshitz} dressed with scalars.
Next we present  a thorough discussion of the different scales present in the solution and the hierarchy among them, which depends on the magnitude of the quark density. Finally, we address the low-temperature thermodynamics. In particular, we determine the dependence of the entropy density on the different dimensionful scales.

In section \ref{sec.flows} we show how to construct domain wall solutions interpolating between the scaling solutions in the IR and the D$p$-brane solution in the UV, showing that both phases are connected by an RG flow, and discuss some dual configurations of the solutions of section \ref{sec.solutions}.

Section \ref{sec.other} explores other setups in which an IR solution with a metric of the form \eqref{eq.genericHVLifshitz} appears.

We finish  in Section \ref{sec.discussion} discussing what other systems may posses phases similar to the ones discussed in this paper, and presenting some future work.

In the Appendix we give technical details on how to write a truncated action describing 10D supergravity in the presence of the string sources.

\section{The gravity dual of super Yang-Mills theory -- a brief review}
\label{review}
In this section we briefly review the results of Ref.~\cite{Itzhaki:1998dd}. 
Maximally supersymmetric SYM  in $d=p+1$ dimensions with gauge group $SU(\nc)$ can be realized as the worldvolume theory of $\nc$ coincident D$p$-branes in type II string theory in the decoupling limit in which $\ls\to 0$ while the Yang-Mills coupling 
\be
\gym^2 = 2\pi g_s  \, ( 2\pi\ell_s)^{p-3} 
\ee
is kept fixed.  In this limit the dual metric, dilaton and Ramond-Ramond (RR) form read
\bse
\label{eq.DpbranesDavid}
\bal
\d s^2  &= \left( \frac{u}{L} \right)^\frac{7-p}{2} \Big[ - f(u)\, \d t^2 +  \d x_p^2 \Big] +  \left( \frac{L}{u} \right)^\frac{7-p}{2} \frac{\d u^2}{f(u)} + 
\left( \frac{L}{u} \right)^\frac{7-p}{2} u^2 \, \d \Omega_{8-p}^2 \,, 
\label{UVmetric} \\[2.5mm]
 e^{\phi} & =  \left( \frac{u}{L} \right)^\frac{(p-3)(7-p)}{4} \,,
 \label{UVdilatonDavid} \\[3.2mm]
F_{8-p} &= (7-p)\, L^{7-p} \, \omega_{8-p}  \,,
\label{eq.F8minusp}
\end{align}
\ese
with $\d \Omega_n$, $\omega_n$ and $V_n = \int \omega_n$ the 
line element, the volume form and the volume of a unit $n$-sphere, respectively.\footnote{Although we are referring to the transverse manifold as a sphere, our results below will be valid for any compact $(8-p)$-dimensional Einstein manifold as long as additional fluxes are not included.  A change in the  transverse manifold implies a change in the dual field theory being described. In the last section we will comment on other cases.} 
The dimensionful constant $L$ is arbitrary in supergravity, but in the quantum theory its value is fixed to 
\be
L^{7-p} = \frac{ (2\pi \ell_s)^{7-p} }{(7-p)V_{8-p}}\,   g_s N 
\label{L}
\ee
by the quantization condition for D$p$-brane charge
\be
{\int F_{8-p} = 2\kappa_{10}^2\,T_p \, N} \,,
\ee
where the ten-dimensional Newton's constant and the D$p$-brane tension are given by 
\be\label{quantcond}
\frac{1}{2\kappa_{10}^2} = \frac{2\pi}{(2\pi \ell_s)^8 \, g_s^2} \ , \qquad T_{p} = \frac{1}{(2\pi \ell_s)^p\, g_s\, \ell_s}\,.
\ee
Note that, without loss of generality, we work with a dilaton normalized in a $g_s$-independent way, so that the local string coupling is actually $g_s e^\phi$. Consistently, we have included explicit factors of $g_s$ in equations \eqq{quantcond}. Finally, for future reference  we have also included in the metric the blackening factor 
\be
f=1-\left( \frac{\uh}{u} \right)^{7-p}
\ee
necessary to describe the finite-temperature physics of the gauge theory. 

In order to make contact with the RG flow in the gauge theory, it is convenient to introduce a new radial coordinate with units of energy defined through ${U=u/\ell_s^2}$. Roughly speaking, one may identify  $U$ with the energy scale in the gauge theory \cite{Itzhaki:1998dd}.\footnote{Although in this paper we will adopt this identification, we emphasize that, except in the vicinity of an RG fixed point, there is no canonical or unique map between the radial coordinate on the gravity side and the energy scale on gauge theory side (near a fixed point the dilatation symmetry can be used to define such a map).} Thus the UV and the IR regimes of the gauge theory get mapped to the regions with large and small values of $U$, respectively. At a given energy scale $U$, the effective  dimensionless coupling in the SYM theory is then 
\be
\leff^2 \sim \lambda\, U^{p-3} \,,
\label{eff}
\ee
with 
\be
\lambda=\gym^2 \nc
\label{lambda}
\ee
the 't Hooft coupling. The effective coupling becomes of order unity at an energy scale 
\be
\up = \lambda^\frac{1}{3-p} \,.
\ee
The perturbative field theory description is applicable provided $\leff \ll 1$.
In terms of $\leff$, the string coupling and the ten-dimensional curvature scale as
\be
g_s e^\phi \sim \frac{\leff^{(7-p)/2}}{\nc}
\sac \ell_s^2 {\cal R}_{10} \sim \frac{1}{\leff} \,.
\ee
The conditions that both the ten-dimensional curvature in the string frame and the string coupling  be small thus translate into
\be
1\ll \leff^2 \ll \nc^\frac{4}{7-p} \,.
\ee

SYM theories with $p<3$ are superrenormalizable and asymptotically free. At energies $U \gg \up$ a perturbative field theory description is possible. In this region the dilaton goes to zero but the curvature grows, meaning that the supergravity description is not valid. 
In the IR, these theories become strongly coupled. In this region  
the curvature in string units becomes small but the dilaton becomes large.  

In contrast, SYM theories with $p>3$ are IR-free and non-renormalizable. 
At energies \mbox{$U \ll \up$} a perturbative field theory description is possible, whereas the supergravity description is not reliable due to the large curvature. In the UV the gauge theory becomes strongly coupled and a supergravity description becomes applicable, albeit a dual one since the dilaton becomes large. 

The case $p=3$ is special, since the gauge theory is the conformally-invariant ${\cal N}=4$ SYM \cite{Maldacena:1997re}.

Since additional scales will appear when we add a quark density, it is important to keep in mind the scales that are already present in the RG flows for SYM without a quark density. We therefore summarise these  here for each dimension.

\subsection{$d=1$ super Yang-Mills}

For $p=0$ there are three relevant scales, listed in decreasing order: 
\bse\bal
U_\mt{pert} &= \gym^{2/3} N^{1/3}=\lambda^{1/3} \,, \\[1mm]
U_\mt{dual} &= \gym^{2/3} \nc^{1/7}= \lambda^{1/3} \nc^{-4/21} \,, \\[1mm]
U'_\mt{dual} &= \gym^{2/3} \nc^{1/9} = \lambda^{1/3} \nc^{-2/9} \,, 
\end{align}\ese
A perturbative SYM description is valid provided that $U\gg U_\mt{pert}$. In this region the curvature is large so a supergravity description is not reliable. Type IIA supergravity provides an accurate description in the region $U_\mt{dual} \ll U \ll U_\mt{pert}$, where the dilaton and the curvature in string units are small. In the range $U'_\mt{dual} \ll U \ll U_\mt{dual}$, the physics is reliably  described by the uplift of the D0-brane solution to M-theory, namely by a gravitational wave in eleven dimensions. At energies  $U< U'_\mt{dual}$ the correct description involves matrix theory \cite{Banks:1996vh}.

\subsection{$d=2$ super Yang-Mills}

For $p=1$ there are three relevant scales, listed in decreasing order: 
\bse\bal
U_\mt{pert} &= \gym N^{1/2}=\lambda^{1/2} \,, \\[1mm]
U_\mt{dual} &= \gym \nc^{1/6}= \lambda^{1/2} \nc^{-1/3} \,, \\[1mm]
U'_\mt{dual} &= \gym = \lambda^{1/2} \nc^{-1/2} \,, 
\end{align}\ese
At energies above $U_\mt{pert}$ perturbative SYM is applicable. In the region \mbox{$U_\mt{dual}\ll U \ll U_\mt{pert}$}  the  reliable description is given by the D1-brane solution of Type IIB supergravity, in which both the dilaton and the curvature in string units are small. Below $U_\mt{dual}$ the IIB dilaton grows large and one must resort to the S-dual description in terms of the near-core region of IIA fundamental strings. This description is valid in the region $U'_\mt{dual} \ll U \ll U_\mt{dual}$. Finally, below $U'_\mt{dual}$ the curvature of the IIA solution grows large and the appropriate description is provided by a free orbifold conformal field theory.

\subsection{$d=3$ super Yang-Mills}

For $p=2$ there are three relevant scales, listed in decreasing order: 
\bse\bal
U_\mt{pert} &= \gym^2 \nc = \lambda \,, \\[1mm]
U_\mt{dual} &= \gym^2 N^{1/5}=\lambda \,\nc^{-4/5} \,, \\[1mm]
U'_\mt{dual} &= \gym^2 = \lambda \, N^{-1}\,, 
\end{align}\ese
A perturbative SYM description is valid provided that $U\gg U_\mt{pert}$. In this region the curvature is large so supergravity is not reliable. Type IIA supergravity provides an accurate description in the range $U_\mt{dual} \ll U \ll U_\mt{pert}$, where the dilaton and the curvature in string units are both small. In the range $U'_\mt{dual} \ll U \ll U_\mt{dual}$, the physics is reliably  described by an eleven-dimensional solution corresponding to M2-branes delocalized along the M-theory circle. Finally, for \mbox{$U< U'_\mt{dual}$} the correct description is provided by the eleven-dimensional $AdS_4 \times S^7$ solution sourced by M2-branes localized on the M-theory circle.

\subsection{$d=4$ super Yang-Mills}

In this case the gauge theory is conformal and therefore there are  no scales. The IIB supergravity description is valid provided $\lambda \gg 1$ and $\nc \gg 1$.

\subsection{$d=5$ super Yang-Mills}

For $p=4$ there are two relevant scales, listed in decreasing order: 
\bse\bal
U_\mt{dual} &= \gym^{-2} N^{1/3}=\lambda^{-1} \nc^{4/3} \,, \\[1mm]
U_\mt{pert} &= \gym^{-2} \nc^{-1}= \lambda^{-1} \,.
\end{align}\ese
A perturbative SYM description is valid provided that $U\ll U_\mt{pert}$. In
 the intermediate regime \mbox{$U_\mt{pert} \ll U \ll U_\mt{dual}$} the dilaton and the ten-dimensional curvature in string units are small, so  Type IIA supergravity provides a reliable description. In the region $U \gg U_\mt{dual}$ the dilaton is large and the correct description is obtained by uplifting the solution to M-theory, which yields $AdS_7 \times S^4$ with one of the spatial $AdS$ directions compactified on a circle. The dual theory is the 6-dimensional M5-brane worldvolume theory, the (0,2) CFT,  compactified on a circle, which provides a UV completion of $d=5$ SYM theory.

\subsection{$d=6$ super Yang-Mills}

Although we will include $p=5$   in our analysis for completeness, we recall that in this case there is no complete decoupling between a `near-brane region' and the asymptotically flat region \cite{Itzhaki:1998dd}. There are two relevant scales, listed in decreasing order: 
\bse\bal
U_\mt{dual} &= \gym^{-1} N^{1/2}=\lambda^{-1/2} \nc \,, \\[1mm]
U_\mt{pert} &= \gym^{-1} \nc^{-1/2}= \lambda^{-1/2} \,.
\end{align}\ese
At energies much below $U_\mt{pert}$ the effective SYM coupling is small and the perturbative description is reliable. At energies $U_\mt{pert} \ll U \ll U_\mt{dual}$ both  the dilaton and the ten-dimensional curvature in string units are small, so the D5-brane solution of IIB supergravity provides a valid description. At energies $U\gg U_\mt{dual}$ the dilaton becomes large and the reliable description is provided by the S-dual solution of IIB supergravity sourced by $\nc$ NS5-branes.

\section{Adding a heavy-quark density to super Yang-Mills}
\label{sec.solutions}
\addtocontents{toc}{\protect\setcounter{tocdepth}{2}}

We will now study the SYM theories above in the presence of a homogeneous density $N_q$ of infinitely-heavy, non-dynamical quarks. On the gravity side the quarks are represented by strings stretching along the radial direction from the bottom of the geometry to the boundary. As the quarks, the strings are homogeneously distributed along the $p$ spatial directions of the gauge theory. Moreover, as explained in the Introduction, we also distribute (smear) the strings homogeneously along the directions of the compact manifold of the ten-dimensional geometry. In order to take into consideration the presence of the strings, we  modify the (string-frame) supergravity action by adding a string source as follows:
\be
S_\mt{total} = S_\mt{IIA/IIB} - \frac{N_q}{2\pi\ell_s^2} \int \left( \sqrt{-G_{tt}\,G_{rr}} \, \d t \wedge \d r - B_2 \right) \wedge \Xi_8 \,,
\label{total}
\ee
where as usual $1/2\pi\ell_s^2$ is the string tension. The so-called `smearing form' 
\be
\Xi_8 = \frac{1}{V_{8-p}}\,  \d x^1 \wedge \cdots \wedge \d x^p \wedge \omega_{8-p} 
\ee
allows us to turn the two-dimensional worldsheet action of the strings into a ten-dimensional integral over the entire spacetime. The
fact that it has `legs' along the gauge theory spatial directions and along the compact directions reflects the smearing of the strings discussed above. Moreover, the normalization with the inverse volume factor in front ensures that the constant $N_q$ appearing in \eqq{total} is exactly the density of strings, or equivalently the density of quarks, per unit volume in the gauge theory spatial directions, with dimensions ${[N_q]=\text{length}^{-p}}$. Henceforth we will refer to $N_q$ indistinctively as  the `string density' or the `quark density'.

The string source in \eqq{total} contributes to the equation of motion of the supergravity Neveu-Schwarz (NS) three-form $H$ which, depending on whether we consider Type IIA or IIB supergravity, reads\footnote{Since we will be interested in solutions with $H=0$, in this equation we have chosen to omit all the terms that vanish when $H=0$ except for the one coming from the kinetic term for $H$.}   
\bse\label{eq.Heqs}
\bal
\d \left( e^{-2\phi} * H \right) + F_2 \wedge  F_6 - \frac{1}{2} F_4 \wedge F_4 & = - \frac{2\kappa_{10}^2}{2\pi\ell_s^2} \, N_q \, \Xi_8 \ ,
\qquad \mbox{[Type IIA]} \\
\d \left( e^{-2\phi} * H \right) + F_1 \wedge F_7 - F_3 \wedge F_5 & = - \frac{2\kappa_{10}^2}{2\pi\ell_s^2} \, \, N_q \, \Xi_8 \ , 
\qquad \mbox{[Type IIB]} 
\end{align}
\ese
with $F_7=-*F_3$ and $F_6 = - * F_4$. Following \cite{Kumar:2012ui}, we note that a simple way to solve these equations is to 
set\footnote{Although it would be interesting to find more general solutions with non-zero $H$, this is beyond the scope of this work.} $H=0$ and to turn on an appropriate RR field strength of the form 
\be\label{eq.Fpform}
F_p = (-1)^{\left[ \frac{p+1}{2} \right]} \frac{{Q}}{L}\, \d x^1 \wedge \cdots \wedge \d x^p \ ,
\ee
where $[x]$ stands for the integer part of $x$ and the dimensionless constant $Q$ is given by\footnote{Although we use the same notation as in \cite{Kumar:2012ui}, our definition of $Q$ is not exactly analogous to that in \cite{Kumar:2012ui} since we have defined it to be dimensionless.}
\be
Q \, = \, \frac{N_q \, 2\kappa_{10}^2}
{2\pi\ell_s^2 \, (7-p) \, V_{8-p} \,  \, L^{6-p}} \, \propto \,  
\ell_s^{4\frac{6-p}{7-p}} \, \lambda^\frac{8-p}{7-p} \, \frac{N_q}{N^2}\,,
\label{Q}
\ee
where in the last equation we have omitted a purely numerical factor. Note that we  assume that $N_q$ and $Q$ are non-negative without loss of generality. If anti-strings are considered there is a physically equivalent solution with positive $Q$ that is obtained by adding an extra minus sign on the right-hand side of Eqs.~\eqq{eq.Heqs} and \eqq{eq.Fpform}.

$F_p$ preserves the rotational invariance in the gauge theory spatial directions and, through its product with $F_{8-p}$ given in \eqq{eq.F8minusp}, `soaks up' the contribution sourced by the strings, thus allowing us to keep $H=0$. Following \cite{Kumar:2012ui} we notice that the RR flux \eqq{eq.Fpform} suggests the presence of dissolved $(8-p)$-baryonic branes wrapped around the $(8-p)$-sphere, each of them obtained by binding $\nc$ quarks into a baryon.

As suggested by \eqq{eq.Fpform}, $Q$ is related to the backreaction of the strings on the D$p$-brane geometry or, equivalently, of the heavy quark density on the SYM dynamics. We will come back to this point below, where we will show that this backreaction depends on the energy scale at which the theory is probed. For the moment we note that in most of this paper the quark density will appear divided by $\nc^2$. We will therefore work with a normalized quark density defined through 
\be\label{eq.nqdefinition}
\nq \equiv \frac{N_q}{N^2} \,.
\ee

Setting to zero all the RR forms except for $F_p$ and $F_{8-p}$ solves their corresponding equations of motion. In order to solve the Einstein  equations, we adopt the following  ansatz for the ten-dimensional metric in Einstein frame 
\be
\d s_{10}^2  = e^{-\frac{2(8-p)}{p}\eta} \,\d s_{p+2}^2 + e^{2\eta} \,L^2 \,\d\Omega_{8-p}^2  \,.
\ee
This preserves the symmetries along the compact directions because we assume that the scalar field $\eta$ depends only on the $p+2$ non-compact directions. Instead of working with the ten-dimensional equations, it is convenient to perform a dimensional reduction along the internal directions to obtain an effective action in $p+2$ dimensions in terms of the lower-dimensional metric $g$, the scalar $\eta$ and the dilaton $\phi$.\footnote{We have corroborated the validity of the truncation by directly obtaining the same solutions from 10D.} The result is (see the Appendix) 
\be\label{eq.pplus2daction}
S_Q  = \frac{1}{2\kappa_{p+2}^2}  \int \d x^{p+2} \, \sqrt{-g} \left[ {\cal R}[g] - \frac{8(8-p)}{p} \, \partial_\mu \eta \partial^\mu \eta \, - \frac{1}{2} \partial_\mu \phi \partial^\mu \phi - V_Q(\eta,\phi) \right] \ ,
\ee
where the potential takes the form
\be\label{eq.potential}
V_Q  = \frac{1}{2\,L^2} \left( \frac{Q}{g_{xx}^{p/2}} \, e^{\frac{5-p}{4}\phi}\, e^{\frac{(8-p)(p-1)}{p}\eta}  +  (7-p) \, e^{\frac{p-3}{4}\phi} \, e^{\frac{(p-8)(p+1)}{p} \eta} \right)^2 - \frac{(8-p) \, (7-p)}{L^2} e^{-\frac{16}{p} \eta}  \ .
\ee
Although we have not made it explicit in \eqq{eq.pplus2daction}, we  assume that the $(p+2)$-dimensional metric is rotationally invariant along the gauge theory spatial directions, and hence that it takes the form 
\be
\d s_{p+2}^2 = g_{tt} \, \d t^2 + g_{xx} \, \d x_p^2 + g_{rr}\, \d r^2 \,.
\ee
The first term inside the brackets in the potential depends on the component of the metric along these spatial directions, $g_{xx}$. The fact that this term is not generally covariant is expected, since the quark density breaks the symmetry between time and space. The term quadratic in $Q$ in the potential comes from the $F_p\wedge*F_p$ term in the supergravity action, whereas the term linear in $Q$ comes from the Nambu-Goto action of the strings. The subindex `$Q$' in the action and in the potential is just a reminder that these quantities  depend explicitly on the quark density.

The dilaton field is dual to a marginal operator ${\cal O}_{p+1}$,  and the breathing mode (volume) of the sphere, $\eta$,  to an irrelevant operator of dimension $\Delta=2(p+1)$,  ${\cal O}_{2(p+1)}$ .
To avoid deforming the Lagrangian of the gauge theory with an explicit inclusion of this irrelevant operator we restrict ourselves to solutions in which the non-normalizable mode of $\eta$ vanishes.

\subsection{UV asymptotics}
In the absence of strings ($\nq={Q}=0$) the D$p$-brane solution \eqq{eq.DpbranesDavid} is recovered as the following solution of \eqref{eq.pplus2daction}: 
\bse\label{eq.Dpbranes}
\bal
\d s_{p+2}^2 & = \left(  \frac{u}{L} \right)^\frac{9-p}{p} \Big[ - f(u)\, \d t^2 +  \d x_p^2 \Big] +  \left( \frac{u}{L} \right)^\frac{(p-4)^2-7}{p} \frac{\d u^2}{f(u)} \,, 
\label{UVmetricc} \\[1mm]
 e^{\phi} & = \left( \frac{u}{L} \right)^\frac{(p-3)(7-p)}{4} \,, 
 \label{UVdilaton} \\[2mm]
\eta &= \frac{p-3}{4(7-p)} \phi \,.
\end{align}
\ese
Note that $\phi$ and $\eta$ are proportional to one another. 
The asymptotic solution in the UV,  $u\to \infty$, for $p\leq5$ is now given by
\bse\label{eq.UVexpansion}
\bal
g_{uu} & =  \left( \frac{u}{L} \right)^\frac{(p-4)^2-7}{p} \ , \\[2mm]
g_{tt} & = - \left( \frac{u}{L} \right)^\frac{9-p}{p} \left[ 1 - \alpha_t \,{Q}\, \left(\frac{L}{u}\right)^{6-p} - \varepsilon \left(\frac{L}{u}\right)^{7-p} + {\cal O} \left(\frac{L}{u}\right)^{2(6-p)} \right] \ , \\[2mm]
g_{xx} & = \left( \frac{u}{L} \right)^\frac{9-p}{p} \left[ 1+\alpha_x \,{Q}\, \left(\frac{L}{u}\right)^{6-p} + P \left(\frac{L}{u}\right)^{7-p} + {\cal O} \left(\frac{L}{u}\right)^{2(6-p)}  \right] \ , \\[2mm]
e^{\phi} & =  \left( \frac{u}{L} \right)^\frac{(p-3)(7-p)}{4} \left[ 1 - \alpha_\phi \,{Q}\, \left(\frac{L}{u}\right)^{6-p} + v_\phi \left(\frac{L}{u}\right)^{7-p} + {\cal O} \left(\frac{L}{u}\right)^{2(6-p)}  \right] \ , \\[2mm]
e^{\eta} & = \left( \frac{u}{L} \right)^\frac{(p-3)^2}{16} \left[ 1 + \alpha_\eta \,{Q}\, \left(\frac{L}{u}\right)^{6-p} + \gamma \left(\frac{L}{u}\right)^{7-p} + {\cal O} \left(\frac{L}{u}\right)^{2(6-p)}  \right] \ , 
\end{align}
\ese
where 
\bse\label{eq.pressureandgamma}
\bal
P&=\frac{1}{p} \varepsilon + \frac{ 8\, (p-3)}{p^2\, (7-p) } v_\phi - \delta_{p,5} \, \frac{107}{1800}\,  {Q}^2 \,, \\[2mm]
\gamma&= \frac{p-3 }{4(7-p) } \, v_\phi- \delta_{p,5} \, \frac{247}{864}\, {Q}^2 \ ,
\end{align}
\ese
and the coefficients $\alpha_t, \alpha_x, \alpha_\phi, \alpha_\eta$ are given in Table \ref{tab.UVexpansion}.  We have grouped under the expression 
${\cal O} \left( L/u \right)^{2(6-p)}$ 
terms proportional to $ u^{ - 2 ( 6 - p ) } $ and to $ u^{ - 2 ( 6 - p ) } \log[u] $.
The form of this expansion shows that the presence of the strings only induces subleading corrections to the UV form of the solution.

\begin{table}[t]
\centering
\begin{tabular}{c|cccc}
$p$ &  $\alpha_t$ & $\alpha_x$ & $\alpha_\phi$ & $\alpha_\eta$ \\[1mm]
\hline\\[-3mm]
$1$ & $\rfrac{-6}{5}$ & $\rfrac{12}{5}$ & $\rfrac{12}{11}$ & $\rfrac{15}{154}$ \\[2mm]
$2$ & $\rfrac{1}{2}$ & $\rfrac{3}{4}$ & $\rfrac{25}{36}$ & $\rfrac{19}{432}$ \\[2mm]
$3$ & $\rfrac{8}{9}$ & $\rfrac{4}{9}$ & $\rfrac{2}{3}$ & $\rfrac{1}{70}$ \\[2mm]
$4$ & $\rfrac{6}{5}$ & $\rfrac{3}{10}$ & $\rfrac{3}{4}$ & $\rfrac{-3}{80}$ \\[2mm]
$5$ & $2$ & $0$ & $\rfrac{4}{3}$ & $\rfrac{-5}{18}$ 
\end{tabular}
\caption{\small The $p$-dependent coefficients appearing in the asymptotic solution \eqref{eq.UVexpansion}.}\label{tab.UVexpansion}
\end{table}

We see that the asymptotic form of the equations of motion does not fix the two coefficients $\varepsilon$ and $v_\phi$. This is expected, since they are  related to the VEVs of the stress energy tensor (specifically to the energy density) and of ${\cal O}_{p+1}$, respectively.
The quantity $P$ is related to the pressure of the system, and receives contributions from $v_\phi$.
There is a third undetermined parameter, $v_\eta$, related to the VEV of the operator ${\cal O}_{2(p+1)}$.
Given that the dimension of this operator is twice the dimension of the stress-energy tensor and ${\cal O}_{p+1}$ it is clear that the order at which the VEV $v_\eta$ would enter the asymptotic expansion would be ${\cal O}(u^{2(p-7)})$, and for this reason it is not displayed.

The term $\delta_{p,5}$ in \eqref{eq.pressureandgamma} is simple to understand.
The contributions from ${Q}$ to the asymptotic expansion occur in powers of $u^{p-6}$, while the VEVs for the operators dual to the metric and the dilaton enter as $u^{p-7}$ powers.
For $p=5$ the quadratic term in $Q$ enters precisely at the same order as the VEVs $\varepsilon$ and $v_\phi$, hence the ${Q}^2$ term in \eqref{eq.pressureandgamma}.

One final conclusion from  \eqref{eq.UVexpansion} is that for $p\geq6$ the $Q$-dependent terms modify the leading terms in the expansion, and hence the inclusion of the quarks cannot be viewed as simply modifying the state in the original SYM theory but as modifying the theory itself.

\subsection{IR solutions}
\label{IRsolutions}

In contrast to the previous section, we will now see that the inclusion of a quark density completely changes the IR properties of the geometry. Here we will  find exact solutions that describe the deep IR regime, and in section \ref{sec.flows}  we will show that they are the endpoints of RG flows that start with the UV asymptotic solutions \eqq{eq.UVexpansion}. We need to distinguish between $d\neq5$ and $d=5$, although as we will see the $d=5$ case can be understood as a smooth  limit of the $d\neq 5$ case. 

\subsubsection{$d\neq 5$} 
The equations of motion derived from \eqref{eq.pplus2daction} admit the following exact solution, to which we will refer as Hyperscaling-violating-Lifshitz (HV-Lif):
\bse\label{IRgeneric}
\bal
\d s_{p+2}^2 &=  \left( \frac{r}{L} \right)^{-\frac{2\,\theta}{p}} \left[ - \left( \frac{r}{L}\right)^{2\, z} f(r)\, \d t^2 + \left( \frac{r}{L}\right)^{2} \d x_p^2 + \beta_\ell^2\,  {Q}^{\frac{2\,(3-p)}{p}}\, \left( \frac{L}{r}\right)^{2} \frac{\d r^2}{f(r)} \right] \,, \label{eq.genericscalinginit} \\[2mm]
e^\phi  &= \beta_\phi\,  {Q}^{\frac{p-7}{2}} \left( \frac{r}{L} \right)^\frac{p\,(p-7)}{2\,(p-4)} \ , \qquad
e^{2\eta}  = \beta_\eta \, {Q}^{\frac{3-p}{4}} \left( \frac{r}{L} \right)^\frac{p\, (3-p)}{4\,(p-4)} \,. \label{eq.IRscalars}
\end{align}
\ese
The dynamical and hyperscaling violating exponents are given by
\be\label{eq.genericscalingend}
z = \frac{16-3\, p}{4-p} \ , \qquad \theta = \frac{p\, (3-p)}{4-p}  \,,
\ee
and the different coefficients $\beta_\eta, \beta_\phi, \beta_\ell$ appearing in the scalars and the metric are listed in Table \ref{tab.uglyfactors}.
The blackening factor is 
\be \label{eq.blackening}
f(r)  = 1 -  \left(\frac{ \rh }{ r } \right)^{p-\theta+z}  \ .
\ee
For future reference, we note that at zero temperature the ten-dimensional Ricci scalar for the string-frame metric reads
\be
{\cal R}_{\rm 10,string} \propto - \frac{ Q}{ L^2 } 
\left(  \frac{L}{r} \right)^\frac{p}{4-p} \,,
\label{10Dcurvature}
\ee
where the proportionality factor is a $p$-dependent positive number.

\begin{table}[t]
\centering
\begin{tabular}{c|ccc}
$p$ &  $\beta_\eta$ & $\beta_\phi$ & $\beta_\ell$ \\[1mm]
\hline\\[-3mm]
$1$ & $\frac{9}{17} \,\sqrt{\frac{6}{5}}$ & $\frac{17^4 \cdot 5^3}{2\cdot 3^{11}}$ & $\frac{2^3 \cdot 3^8}{5^2 \cdot 17^4} \sqrt{21}$ \\[2mm]
$2$ & $\frac{5^{1/4} \, \sqrt{11}}{7^{3/4} \, \sqrt{2}}$ & $ \frac{2^5 \cdot 7^3}{11^3} \, \sqrt{\frac{7}{5}}$ & $\frac{11}{7} \, \sqrt{\frac{3}{7}}$ \\[2mm]
$3$ & $\frac{2^{3/4} \cdot 17^{1/4}}{\sqrt{11}}$ & $ \frac{11^3 \cdot 3^2}{17^2 \cdot 2 \, \sqrt{34}}$ & $ \frac{2 \cdot 17^{1/3}}{11^{2/3}} \, \sqrt{10} $ \\[2mm]
$4$ & $\frac{3^{1/4}}{2^{1/4}}$ & $\frac{16\, \sqrt{2}}{9\, \sqrt{3}}$ & $\frac{2^{1/4}}{3^{1/4}}$ \\[2mm]
$5$ & $\frac{\sqrt{2}}{7^{1/4}}$ & $\frac{5^2}{2 \cdot 7\, \sqrt{7}}$ & $\frac{2^{7/5}}{7^{1/5}} \sqrt{3}$ 
\end{tabular}
\caption{\small The $p$-dependent factors appearing in the IR solution \eqref{IRgeneric}.}\label{tab.uglyfactors}
\end{table}

Under the transformation \eqref{eq.Lifshitzscaling}, together with $r \to \Lambda^{-1} \, r $, the  metric \eqq{eq.genericscalinginit} transforms as \eqq{as}. This scaling property is mildly broken by the logarithmically-running scalars \eqq{eq.IRscalars}.

The  solution \eqref{IRgeneric} satisfies the null energy conditions \eqref{eq.NECs}, even for $p=5$ with a negative dynamical exponent $z=-1$.
When $p=6,7,8$ there is no HV-Lif solution in the IR.
Presumably, this is connected with the observation above that for those high values of $p$ the string density modifies the leading UV behavior of the theory.

\subsubsection{$d=5$}\label{sec.D4strings}
For $p=4$ the system \eqref{eq.pplus2daction} admits the exact solution 
\bse\label{IR4}
\bal
\label{eq.pis4solution}
 \d s_6^2 &= \left(  \frac{r}{L} \right)^{1/2 } \left[ - \left( \frac{r}{L} \right)^2 f(r) \, \d t^2 + \d x_4^2 + \beta_\ell^2\, {Q}^{-1/2} \left( \frac{L}{r} \right)^2 \frac{\d r^2}{f(r)}   \right] \,, \\[2mm]
e^{\phi}  &= \beta_\phi \, {Q}^{-\rfrac{3}{2}}  \left(\frac{r}{L} \right)^\frac{3}{2}\ , \qquad 
e^{2\eta}  = \beta_\eta \, {Q}^{-\rfrac{1}{4}}  \left(\frac{r}{L} \right)^\frac{1}{4}\,,
\label{IRscalars4}
\end{align}
\ese
where the coefficients $\beta_\ell, \beta_\phi, \beta_\eta$ are given in  Table \ref{tab.uglyfactors} and the blackening factor is 
\be
f(r)=1-\left(\frac{ \rh }{ r } \right)^{2} \,. 
\label{black}
\ee
The metric in this solution is conformal to  $AdS_2\times \mathbb{R}^4$ and can be obtained from the generic-$p$ solution \eqref{IRgeneric}--\eqref{eq.genericscalingend}
 via a redefinition of the radial variable 
\be\label{eq.pis4change}
\left(  \frac{r}{L} \right) \to \left(  \frac{r}{L} \right)^{1/z} \ ,
\ee
followed by the $p\to4$ limit and a trivial rescaling of the gauge-theory directions. This limit can be understood as the limit $z\to\infty$ with the ratio $z/\theta=-1$ fixed. For future reference, we note that at zero temperature the ten-dimensional Ricci scalar in the string frame for this solution reads
\be
{\cal R}_{\rm 10,string} = - \frac{189}{8}\, \frac{{Q}}{L^2}\,  \frac{L}{r} \ .
\ee

\subsection{Physical scales and qualitative RG flows}
\label{scales}
Even without knowledge of the entire RG flow, in this section we will be able to anticipate the parametric form of several energy scales that will play an important role in the entire flows. To do so we need  a map between the IR radial coordinate $r$ used in \eqq{IRgeneric} and \eqq{IR4} and the energy scale in the gauge theory. In order to be consistent with the choice we made for this map in the UV, we will actually map the IR radial coordinate $r$ to the UV radial coordinate $U$. The requirement that the string density in the IR solution is the same as in the UV solution, as it should be if they are both connected through an RG flow, translates into the requirement that the $g_{xx}$ components of both metrics be equal when expressed in terms of the same coordinate. This leads to the following identifications:
\bse
\label{rU}
\bal
\label{eq.relationIRradiusenergy}
 p\neq 4:& \,\,\, \,\,\, \,\,\, 
 \frac{r}{L} \sim  \left( \frac{U\, \ell_s^2}{L} \right)^\frac{(9-p)(4-p)}{2p}  \,, 
 \\[2mm]
 p = 4: & \,\,\, \,\,\, \,\,\, 
 \frac{r}{L} \sim \left( \frac{U\, \ell_s^2}{L} \right)^{5/2} \,,
 \label{eq.relationIRradiusenergy4}
\end{align}
\ese
where in the first equation we have used the explicit form of the HV exponent $\theta$ given in 
\eqq{eq.genericscalingend}. Note that for $p<5$ this means that the IR corresponds to small values of $r$, whereas for $p=5$ the IR  corresponds to  large values  of $r$. This observation, together with Eqn.~\eqq{eq.IRscalars}, implies that, in all cases, the dilaton starts off in the deep IR growing towards the UV. In the complete RG flows that we will construct in the next section this behavior will be matched onto the UV behavior \eqq{UVdilaton}. This means that for $p \geq 3$ the dilaton will grow monotonically from the IR to the UV, whereas for $p < 3$ the dilaton reaches a maximum after which it decreases. 

There are three energy scales that will play an important role in all the flows with non-zero quark density. The first one is the crossover scale at which the effect of the quark density becomes of order unity. Looking at the UV expansion \eqq{eq.UVexpansion} we see that the backreaction of the strings scales as
\be
 {Q}\, \left(\frac{L}{u}\right)^{6-p} \sim 
\frac{\nq \, \lambda^2}{U^{6-p}} \,.
\ee
As anticipated above, this is scale-dependent and small at high energies, but it becomes of order unity at the scale 
\be\label{eq.crossover}
\uc 
\sim \lambda^\frac{2}{6-p} \,\nq^\frac{1}{6-p} \ .
\ee
Consistently, at this scale the geometry transitions between the UV and the IR asymptotic regimes, as can be seen by the fact that at $U\sim \uc$ we have 
${g_s \, e^{\phi_\mt{UV}} \sim g_s\, e^{\phi_\mt{IR}}}$, where ${\phi_\mt{UV}}$ is given by \eqq{UVdilaton}  and  ${\phi_\mt{IR}}$ is given  by \eqq{eq.IRscalars} with $r$ replaced by $U$ according to \eqq{rU}.

 The second important scale is the scale at which the IR local string coupling, $g_s e^{\phi_\mt{IR}}$, becomes of order unity, and it is given by 
\be\label{eq.dilatonIR}
\uu 
\sim N^\frac{4}{(7-p)(9-p)} \, \lambda^\frac{3}{9-p} \, \nq^\frac{2}{9-p} \,,
\ee

The third important scale is the scale
\be\label{eq.RicciIR}
\ul 
\sim  \lambda^\frac{3}{9-p} \, \nq^\frac{2}{9-p} 
\ee
above which the curvature of the IR metric \eqq{eq.genericscalinginit} or \eqq{eq.pis4solution} is small in string units. We thus see that the IR solutions \eqq{IRgeneric} or \eqq{IR4} can be trusted in the region $\ul \ll U \ll \uu$.
Note that at large $\nc$ this region is parametrically large, since 
\be
\uu = N^\frac{4}{(7-p)(9-p)} \ul \,.
\ee

The complete RG flows that we will construct in  section \ref{sec.flows} can be roughly understood as taking a piece of the UV solution \eqq{eq.Dpbranes} and a piece of the IR solutions \eqq{IRgeneric} or \eqq{IR4}, and gluing them together along a small transition region at the scale $\uc$. Even without knowledge of the entire flows, we will now show that these can be classified in several qualitatively distinct classes depending on how the quark density $\nq$ compares to other scales in the theory. We need to distinguish between the cases $d=2, 3$, $d=5, 6$ and $d=4$. The case $d=1$ is special and will be discussed in Section \ref{sec.D0branes}.

\subsubsection{$d=2, 3$}\label{sec.plessthan3}
In these cases there are three characteristic densities corresponding to the values of $\nq$ at which $\uc$ becomes of order $\up, \ud$ or $\udp$, respectively. With an obvious notation and listed in decreasing order, these values are:
\bse
\label{densities}
\bal
\nq^\mt{pert}\sim & \,\, \lambda^{\frac{p}{3-p}} \,, \\[2mm]
\nq^\mt{dual} \sim &\,\, \lambda^{\frac{p}{3-p}} \,
\nc^{\frac{4(p-6)}{(p-7)(p-3)}} \,, \\[2mm]
{\nq'}^{\mt{dual}} \sim &\,\, \lambda^{\frac{p}{3-p}} \, 
\nc^{\frac{41p^2 -69p-32}{24}} \,.
\end{align}
\ese
Note that, although the definitions of these densities make no reference to $\ul$ and $\uu$, at $\nq \sim \nq^\mt{pert}$ we have $\uc \sim \up \sim \ul$ and at $\nq \sim \nq^\mt{dual}$ we have \mbox{$\uc \sim \ud \sim \uu$}.
The densities above define several regions in which the following 
  hierarchies of energy scales hold: 
\bse
\label{hierarchy}
\bal
\nq^\mt{pert}  \ll \nq & \quad\rightarrow \quad 
\udp \ll \ud \ll \up \ll \uc \ll \ul \ll \uu  
\label{rg1} \,, \\[2mm]
\nq^\mt{dual} \ll \nq \ll \nq^\mt{pert} & \quad\rightarrow \quad 
\udp \ll \{ \ud \,, \ul\} \ll \uc \ll \{ \uu \,, \up \}  
\label{rg2} \,, \\[2mm]
{\nq'}^{\mt{dual}} \ll \nq \ll \nq^\mt{dual} & \quad\rightarrow \quad 
\{ \udp \,, \ul\}  \ll \uu \ll \uc \ll \ud \ll \up  
\label{rg3} \,, \\[2mm]
\nq \ll {\nq'}^{\mt{dual}} & \quad\rightarrow \quad 
\ul \ll \uu \ll \uc \ll \udp \ll \ud \ll \up  
\label{rg4}  \,.
\end{align}
\ese
To clarify our notation, an expression like $\uc \ll \{ \uu \,, \up \}$ means that 
$\uc$ is much smaller than both scales inside the curly brackets, but the ordering between the latter is not specified. 

\begin{figure}[t]
\begin{center}
\includegraphics[width=0.65\textwidth]{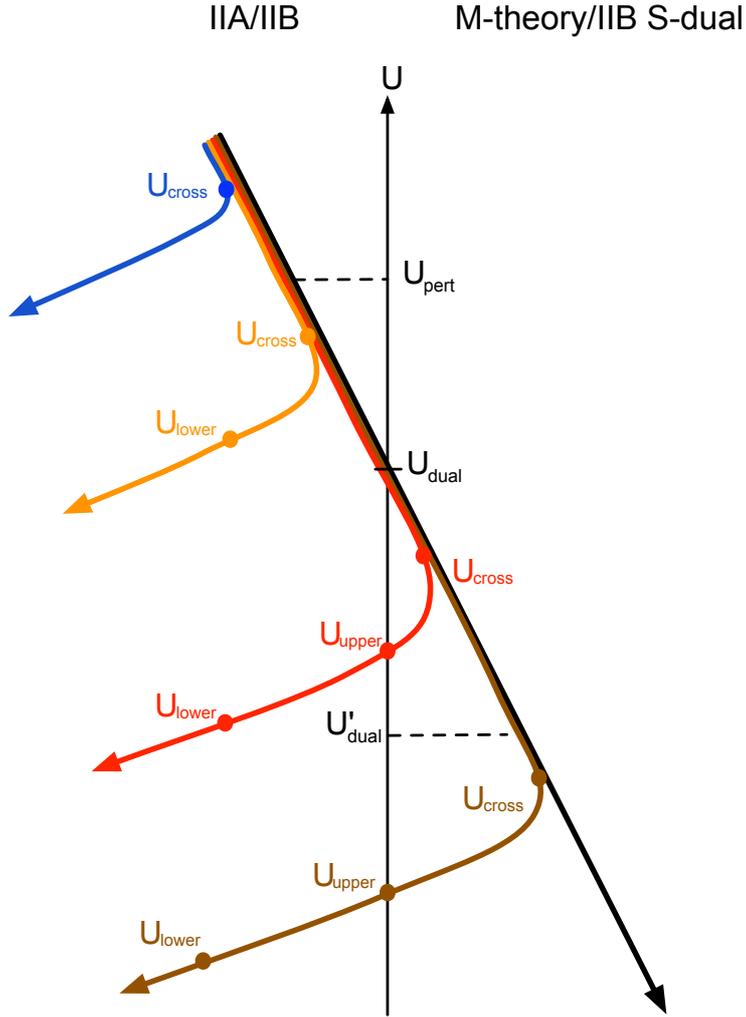}
\caption{\small \label{Diagram_p_smaller_than_3}
Pictorial representation of the different classes of RG flows for SYM with $d=2, 3$ (see the main text).}
\end{center}
\end{figure}

The four density regions \eqq{hierarchy} give rise to four qualitatively different types of RG flows, as depicted in Fig.~\ref{Diagram_p_smaller_than_3}.
Points to the left or to the right of the vertical line in Fig.~\ref{Diagram_p_smaller_than_3} are meant to correspond to small or large values of the dilaton, respectively. The black, diagonal  line represents the RG flow corresponding to SYM with $p<3$ in the absence of a quark density. In this case the dilaton grows monotonically from the UV to the IR and one encounters the three scales  $\up, \ud$, and $\udp$ described in section \ref{review}. The crucial modification introduced by the quark density is that the dilaton is no longer a monotonic function. Instead, as shown in Figs.~\ref{fig.D2RG} and \ref{fig.D1RG} below, the dilaton achieves a maximum at the scale $\uc$ at which the transition between the UV and the IR regimes takes place. For flows with densities in the region \eqq{rg1} this transition takes place above $\up$, where the theory is still described by perturbative SYM. For these flows therefore there is no energy region that is reliably described by supergravity. An example of a flow in this class is represented by the top, blue curve in Fig.~\ref{Diagram_p_smaller_than_3}.

For flows with densities in the region \eqq{rg2} the transition takes place between $\ud$ and $\up$, as illustrated by the second-from-the-top, orange curve in Fig.~\ref{Diagram_p_smaller_than_3}. In this case  the reliable description at the transition point is given by Type IIA  (for $p=2$) or Type IIB supergravity (for $p=1$). In the region $\uc \ll U \ll \up$ above the transition point, the solution is well approximated by the UV solution \eqq{eq.Dpbranes}, whereas in the region $\ul \ll U \ll \uc$ below the transition point the solution is well approximated by the IR solution \eqq{IRgeneric}. The dilaton never becomes large along these flows and thus there is never a need to resort to a dual description.

For flows with densities in the region \eqq{rg3} the transition takes place between $\udp$ and $\ud$, as illustrated by the third-from-the-top, red curve in Fig.~\ref{Diagram_p_smaller_than_3}. At the transition point  supergravity provides a reliable description, but this is not given by the original solutions \eqq{eq.Dpbranes} in the UV and \eqq{IRgeneric} in the IR. Instead, these flows are well described by the UV solution \eqq{eq.Dpbranes} only down to the scale $\ud$. In the region 
$\uc \ll U \ll \ud$ the reliable description is given by the M-theory uplift (for $p=2$) or  by the S-dual (for $p=1$) of the UV solution \eqq{eq.Dpbranes}.\footnote{To perform S-duality on our solutions one must also S-dualize the string sources into D-strings.}
In the region $\uu \ll U \ll \uc$ the reliable description is given by the M-theory uplift (for $p=2$) or by the S-dual (for $p=1$) of the IR solution \eqq{IRgeneric}. We will come back to these dual solutions below. Finally, in the region $\ul \ll U \ll \uu$ the reliable solution is given by \eqq{IRgeneric}.

For flows with densities in the region \eqq{rg4} the transition takes place below $\udp$. The description of these flows all the way down to the scale $\uc$ is essentially the same as in the case of pure SYM without a quark density. In particular,  the physics just above the transition point is reliably described by the localized M2 solution of M-theory for $p=2$ and by a free orbifold CFT for $p=1$. The far IR region $\ul \ll U \ll \uu$ is well described by the supergravity solution \eqq{IRgeneric}. In contrast, most of the intermediate region $\uu \ll U \ll \uc$ is probably not well described by the naive supergravity dual of the solution \eqq{IRgeneric}. For $p=2$ the reason is that one needs to match the M-theory uplift of this  solution, which is delocalized along the M-theory circle, to the localized M2 solution. For $p=1$ the reason is that one would need to match the S-dual solution to the free orbifold description.

\subsubsection{$d=5, 6$}\label{sec.pgreaterthan3}

In these cases  there are two characteristic densities corresponding to the values of $\nq$ at which $\uc$ becomes of order $\up$ or $\ud$, respectively. With an obvious notation and listed in decreasing order, these values are:
\bse
\bal
\nq^\mt{dual} \sim &\,\, \lambda^{\frac{p}{3-p}} \,, \\[2mm]
\nq^\mt{pert}\sim & \,\, \lambda^{\frac{p}{3-p}} \, \nc^{\frac{4(p-6)}{(p-7)(p-3)}} \,.
\end{align}
\ese
Note that, although the definitions of these densities make no reference to $\ul$ and $\uu$, at $\nq \sim \nq^\mt{pert}$ we have $\uc \sim \up \sim \ul$ and at $\nq \sim \nq^\mt{dual}$ we have \mbox{$\uc \sim \ud \sim \uu$}.
The densities above define several regions in which the following 
  hierarchies of energy scales hold: 
\bse
\label{hierarchybis}
\bal\label{rg1bis}
\nq^{\mt{dual}} \ll \nq & \quad\rightarrow \quad 
 \up \ll \{ \ud  \, , \ul \} \ll \uu \ll \uc  \ ,\\[2mm]\label{rg2bis}
\nq^\mt{pert} \ll \nq \ll \nq^\mt{dual} & \quad\rightarrow \quad 
 \{ \up \, , \ul \} \ll \uc \ll \{ \ud \, ,  \uu  \}  
 \,, \\[2mm]\label{rg3bis}
\nq  \ll \nq^\mt{pert} & \quad\rightarrow \quad 
 \ul \ll \uu  \ll \uc \ll \up \ll  \ud \ ,
\end{align}
\ese

\begin{figure}[t]
\begin{center}
\includegraphics[width=0.65\textwidth]{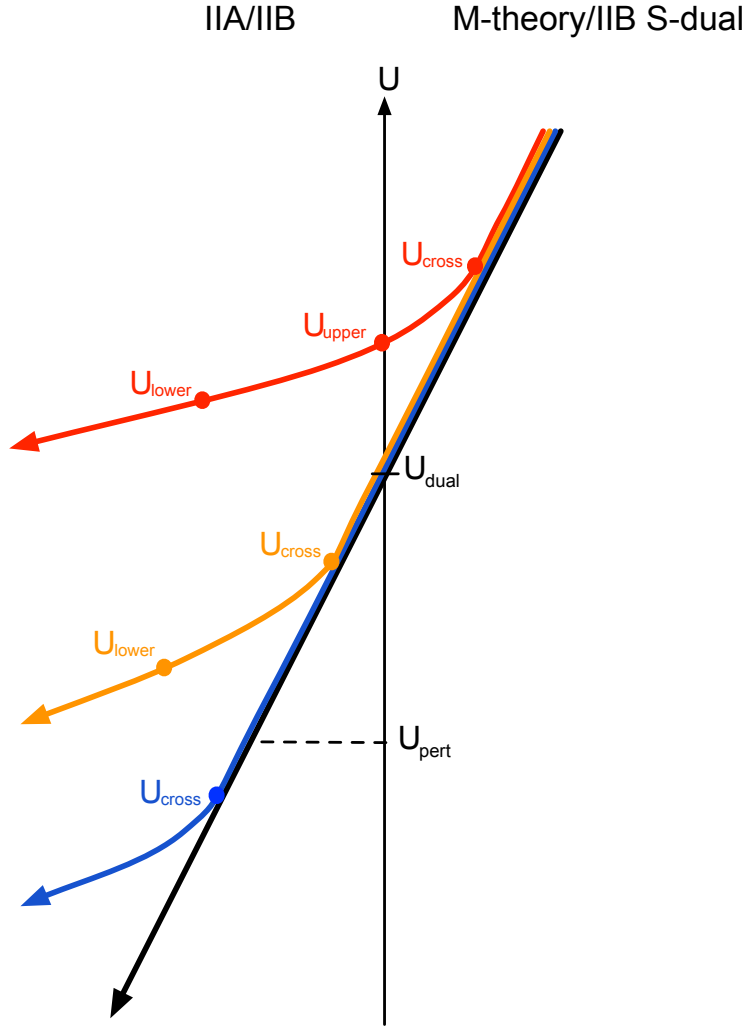}
\caption{\small \label{Diagram_p_larger_than_3}
Pictorial representation of the different classes of RG flows for SYM with $d=5, 6$ (see the main text).}
\end{center}
\end{figure}

The three density regions \eqq{hierarchybis} give rise to three  different classes of RG flows, depicted in Fig.~\ref{Diagram_p_larger_than_3}.

As before, points to the left or to the right of the vertical line in Fig.~\ref{Diagram_p_larger_than_3} correspond to small or large values of the dilaton, respectively. The black, diagonal  line representing the RG flow of $p=4, 5$ SYM in the absence of  quark density shows a monotonic decreasing behavior. The quark density does not change this monotonicity, it just produces a change in the slope. 

For  densities in the region \eqq{rg1bis} the transition between the two behaviors takes place above $\ud$, where the theory is described by the M5-brane solution of M-theory (for $p=4$) or the IIB supergravity solution sourced by NS5-branes (for $p=5$). The far IR region is well captured by the 10D supergravity description of the previous section, but at intermediate scales a dual description is needed, which we will discuss below. The top, red curve in Fig.~\ref{Diagram_p_larger_than_3} depicts an example of this type of flow. 

For flows with densities in the region \eqq{rg2bis} the transition takes place between $\ud$ and $\up$, as illustrated by the second-from-the-top, orange curve in Fig.~\ref{Diagram_p_larger_than_3}. In this case  the reliable description at the transition point is given by Type IIA  (for $p=4$) or Type IIB supergravity (for $p=5$). No dual description is needed to describe the regime where the backreaction of the quark charge is important.

Finally, for flows with densities in the region \eqq{rg3bis} the transition takes place below  $\up$, where the theory is  described by perturbative SYM. For these flows  there is no energy region that is reliably described by supergravity. This is illustrated by the bottom, blue curve in Fig.~\ref{Diagram_p_larger_than_3}.

\subsubsection{$d=4$}\label{sec.pequals3}

In this case $\nq$ is the only scale in the theory and there is only one class of RG flows. Provided the supergravity description is valid in the UV, {i.e.}, that $\lambda\gg1$ and $\nc\gg 1$, we find an energy hierarchy
\be
\ul \ll \uc \ll \up \ .
\ee
The dilaton decreases monotonically (although very slowly in the UV) towards the IR, and the IIB description of the theory is valid down to the scale $\ul$.

\subsection{Low-temperature thermodynamics}\label{sec.thermodynamics}
\addtocontents{toc}{\protect\setcounter{tocdepth}{2}}

In this section we study the thermodynamics of the IR solutions \eqref{IRgeneric} at low temperatures, where the quantitative meaning of `low' will be given below. 
The complete thermodynamics can be derived once the entropy density is known. At low temperatures, the dependence of the entropy density on the temperature $T$ is fixed by a scaling argument to be of the form  ${s\sim T^{(p-\theta)/z}}$, with $\theta$ the hyperscaling-violating exponent and $z$ the  dynamical exponent.
For solutions satisfying the null energy conditions, as the ones obtained in this paper, this gives a positive specific heat, discarding a possible source of instability.
Our goal is to determine the dependence of the entropy density on the other parameters of the theory, namely on $N, \lambda$ and $\nq$. 

The Bekenstein-Hawking formula applied to the metric \eqref{eq.genericscalinginit} or \eqq{eq.pis4solution}, after using the relations \eqref{rU}, gives for the entropy density
\be\label{eq.entropyUT}
s = \frac{2\pi}{\kappa_{p+2}^2} \frac{{\cal A}_{p}}{\int \d^p x} \sim N^2 \, \lambda^{-\frac{3}{2}} \, U_\mt{T}^{\frac{9-p}{2}}  \ ,
\ee
with $U_\mt{T}$ the scale associated to the radius of the horizon (and therefore the temperature) $\rh$.

To obtain the temperature we need to relate  the normalization of the IR time coordinate in \eqref{eq.genericscalinginit} or \eqq{eq.pis4solution} to the normalization of the UV time coordinate in \eqref{eq.Dpbranes}.
These are related by a factor, $t_\mt{UV} = y\, t_\mt{IR}$, which will depend on the parameters of the theory in a specific way.
To calculate this dependence we require that the norm of the Killing vector $\xi=\partial_{t_\mt{UV}}$  is continuous at the crossover scale $\uc$. This norm is given by  the time-time component of the metric, therefore we have to solve
\be
g_{tt}^\mt{UV} (u) \sim  y^{-2}\, g_{tt}^\mt{IR}(u) \bigg|_{u=\uc\, \ell_s^2} \ ,
\ee
where this expression is evaluated at the crossover scale and we have related the IR and UV radii on the r.h.s.~with \eqref{eq.relationIRradiusenergy} (or \eqref{eq.relationIRradiusenergy4} for $p=4$). For simplicity we omit ${\cal O}(1)$ factors, thus obtaining
\be
y \sim Q^{\frac{9-p}{p}} \ .
\ee

The relation between the IR and UV time coordinates enters in the calculation of the temperature via the periodicities of the Euclidean times in both limits of the RG flow
\be
\beta_{\tau}^\mt{UV} \sim Q^\frac{9-p}{p} \beta_{\tau}^\mt{IR} \ .
\ee
This allows us to express the temperature from the requirement of the absence of conical singularities in the Euclidean solution. Using the IR geometry with the appropriately UV-normalized Killing vector $\xi$ gives
\be\label{eq.IRtemperature}
T  \sim \frac{1}{4\pi\, Q^\frac{9-p}{p}} \frac{ - \partial_r g_{tt}^\mt{IR} }{  \sqrt{  - g_{tt}^\mt{IR}   g_{rr}^\mt{IR}  } }\Bigg|_{r\to \rh}   \!\!\!\!\!\!\!\! \sim \lambda^{-\frac{48-7p}{2p}} \, \nq^{-2\frac{6-p}{p}} \, U_\mt{T}^\frac{(9-p)(16-3p)}{2p} \ ,
\ee
which immediately leads to the entropy density 
\be\label{eq.IRentropy}
s \sim  N^2 \, \nq^\frac{2(6-p)}{16-3p} \, \lambda^\frac{p}{16-3p} \, T^\frac{p}{16-3p} \ .
\ee
This expression is valid provided the horizon is in a region where the IR metric provides a reliable description, which depends on how the temperature relates to the charge density. In order to translate from the latter condition on $U_\mt{T}$ to a condition on $T$, let us use \eqq{eq.IRtemperature} to define three temperatures associated to the scales $\ul$, $\uc$ and $\uu$:
\begin{equation}
T_\mt{lower} = \lambda^{-1}\, \nq^\frac{4-p}{p}\ , \qquad T_\mt{cross} =   \lambda^{\frac{4-p}{2(6-p)}} \, \nq^\frac{5-p}{2(6-p)}  \ , \qquad T_\mt{upper} =   \lambda^{-1} \, \nc^\frac{32-6p}{(7-p)\,p} \, \nq^\frac{4-p}{p}  
\,.
\end{equation}
Then we have that \eqref{eq.IRentropy}  is a good approximation to the entropy density for temperatures in the range
\bse
\bal
T_\mt{lower}  \ll T \ll T_\mt{cross} & \quad\text{ for }\begin{cases}
\nq^\mt{dual} \ll \nq  \ll \nq^\mt{pert} & \quad\text{if $p<3$} \ , \\
\text{every }\nq & \quad\text{if $p=3$} \ , \\
\nq^\mt{pert}\ll \nq \ll \nq^\mt{dual} & \quad\text{if $p>3$} \ ,
\end{cases}  \\[2mm]
T_\mt{lower}  \ll T \ll T_\mt{upper} & \quad\text{ for }\begin{cases}
\nq \ll \nq^\mt{dual} & \quad\text{if $p<3$} \ , \\
\nq^\mt{dual} \ll \nq & \quad\text{if $p>3$}  \ .
\end{cases} 
\end{align}
\ese
For $T\ll T_\mt{lower}$ we need a new solution beyond the regime of validity of supergravity, since the Ricci scalar in ten dimensions becomes large for the solutions presented in this paper.

\section{Complete RG flows }\label{sec.flows}
\addtocontents{toc}{\protect\setcounter{tocdepth}{1}}

We will now show that the IR solutions presented in section \ref{IRsolutions} (at zero temperature) are connected to the asymptotic solutions  \eqref{eq.UVexpansion} by RG flows. In other words, we will prove that the IR solutions given above describe the IR physics of SYM theories in the presence of a non-zero density of heavy quarks. Since the gravitational solutions will interpolate between two fixed points, we will refer to them as  `domain wall' solutions. We will first present a general analysis and then solve the corresponding equations numerically on a case-by-case basis. 

We start by using diffeomorphism invariance to fix $g_{xx}(r)$ to its IR functional form, thus fixing the gauge for the radial coordinate.
With this choice we can use Einstein's equation along the radial direction to solve algebraically for $g_{rr}(r)$.
Substituting back in the remaining equations we are left with three  second-order differential equations for $\varphi_i\equiv\left\{g_{tt}(r), \phi(r),\eta(r)\right\}$. 
We will solve these equations numerically integrating from the IR to the UV.
To determine the appropriate boundary conditions in the IR, consider fluctuations around the HV-Lif solution of the form
\be\label{eq.modesansatz}
g_{tt}  = g_{tt}^{(0)} \left( 1 + \delta g_{tt} (r) \right) \ , \quad
e^\phi  = e^{\phi^{(0)}}  \left( 1 + \delta \phi (r) \right) \ , \quad
e^\eta  = e^{\eta^{(0)}} \left( 1 + \delta \eta (r) \right) \ , 
\ee
with the three $\varphi_i^{(0)}$ functions given in \eqref{eq.genericscalinginit} or \eqq{eq.pis4solution}.
The linearised equations of motion for the  fluctuations 
$\delta \varphi_i$ have a general solution given by a linear combination of six independent modes
\be
\label{powers}
\delta \varphi_i = \sum_{j=1}^3 \left( c_{i,j}^{(+)} \, \left( \frac{r}{L} \right)^{\Delta_{+}^{ j}} + c_{i,j}^{(-)}\, \left( \frac{r}{L} \right)^{\Delta_{-}^{ j}} \right)  \,.
\ee
For each of the $\Delta_\pm^j$ modes, one of the $c_{i,j}^{(\pm)}$ is undetermined and the rest (running in $i$) are either proportional to it or vanish.

The six different powers in the fluctuations are organised in three pairs (labelled by $j$), each one satisfying 
$\Delta_+^j + \Delta^j_- = p-\theta+z$. One of these pairs is given by 
$\Delta_{+}^3=p-\theta+z$ and $\Delta_{-}^3=0$ and has  
$\delta g_{tt} (r) =-\delta g_{rr} (r)$. We recognize this as the deformation corresponding to turning on a non-zero temperature. Since we are interested in the RG flow at zero temperature, we will force these modes to vanish by setting $c_{i,3}^{(\pm)}=0$.

At this point we are left with the two pairs corresponding to $j=1$ and $j=2$. For both of them $\Delta_+^j>0$ and $\Delta_-^j<0$.
Integrating from the IR we want to activate just the modes that are regular at $r=0$,\footnote{For $p=5$ the IR is at $r\to\infty$, and the present discussion holds if we take $r\to L^2/r$ first.} corresponding to  perturbations that are irrelevant with respect to the IR fixed point. This implies that we must set $c_{i,j}^{(-)}=0$ as well.

We have thus reduced the IR boundary conditions to a two-dimensional space parametrized by two coefficients that we rename   $c_1$ and $c_2$ for simplicity. Flows that start in the IR with arbitrary values of these parameters generically lead in the UV to a non-zero source of the irrelevant operator of dimension $2(p+1)$ dual to $\eta$. In order to connect with the asymptotic solution \eqref{eq.UVexpansion} we must demand that this source vanishes. This requirement gives a relation between $c_1$ and $c_2$. Any flow determined by a pair $(c_1, c_2)$ that obeys this condition leads to a physical RG flow connecting the correct IR and the correct UV, but in general the normalisation of the dilaton in the UV differs by a constant  from that in \eqq{UVdilaton}. In order to construct the flow with the  normalisation adopted in \eqq{UVdilaton} one must fine-tune the values of both $c_1$ and $c_2$.

It may seem that this procedure must be repeated for each value of the quark density $\nq$, or equivalently for each value of $Q$, but this is not the case because  flows with different values of $Q$ are simply related to one another. To see this we note that upon the rescalings
\be\label{eq.Qredef}
e^\phi \to Q^\frac{p-7}{2} e^\phi \ , \qquad e^\eta \to Q^\frac{3-p}{8} e^\eta \ , \qquad g_{rr} \to Q^\frac{2(3-p)}{p} g_{rr} \ ,
\ee
the effective action \eqref{eq.pplus2daction} and the potential \eqref{eq.potential} transform homogeneously as
\be
V_Q \to Q^\frac{2(p-3)}{p}V_{Q=1} \ , \qquad \quad S_Q \to Q^\frac{p-3}{p} S_{Q=1} \,.
\ee
This means that, given the solution with $Q=1$, the general solution with generic $Q$ is obtained from it simply by rescaling the fields as indicated in \eqref{eq.Qredef}. We will therefore set $Q=1$ in all the domain-wall solutions that we present below.

\subsection{$d=3$ super Yang-Mills}

We begin our discussion with the case $p=2$. The reason is its potential interest of three-dimensional field theories for the description of systems on thin layers. On the same basis, we will provide expression \eqref{eq.expansionIR} for this case only.

The powers introduced in \eqq{powers} that are not related to a temperature deformation take the form 
\be\label{eq.D2-IRmodes}
\Delta_\pm^1  = -3 \pm \sqrt{\frac{3}{11} ( 381 - 4\sqrt{793}) } \ , \qquad
\Delta_\pm^2  = -3 \pm \sqrt{\frac{3}{11} ( 381 + 4\sqrt{793}) } \ .
\ee
The two positive modes are to be identified with the ones in the generic discussion below \eqq{powers}.
The fluctuations of the functions to integrate are
\bse\label{eq.expansionIR}
\bal
-\delta g_{tt} & = c_1 \left( \frac{r}{L} \right)^{\Delta_+^1}  + c_2 \left( \frac{r}{L} \right)^{\Delta_+^2} \ , \\
\delta \phi & = \frac{c_1}{8} \left( 31 - \sqrt{793} \right) \left( \frac{r}{L} \right)^{\Delta_+^1}  + \frac{c_2}{8} \left( 31 + \sqrt{793} \right) \left( \frac{r}{L} \right)^{\Delta_+^2} \ , \\
\delta \eta & = \frac{c_1}{96} \left( 139 - 5 \sqrt{793} \right) \left( \frac{r}{L} \right)^{\Delta_+^1}  + \frac{c_2}{96} \left( 139 + 5 \sqrt{793} \right) \left( \frac{r}{L} \right)^{\Delta_+^2} \ .
\end{align}
\ese
\begin{figure}[t!]
\begin{center}
\begin{subfigure}[b]{0.45\textwidth}
\includegraphics[width=\textwidth]{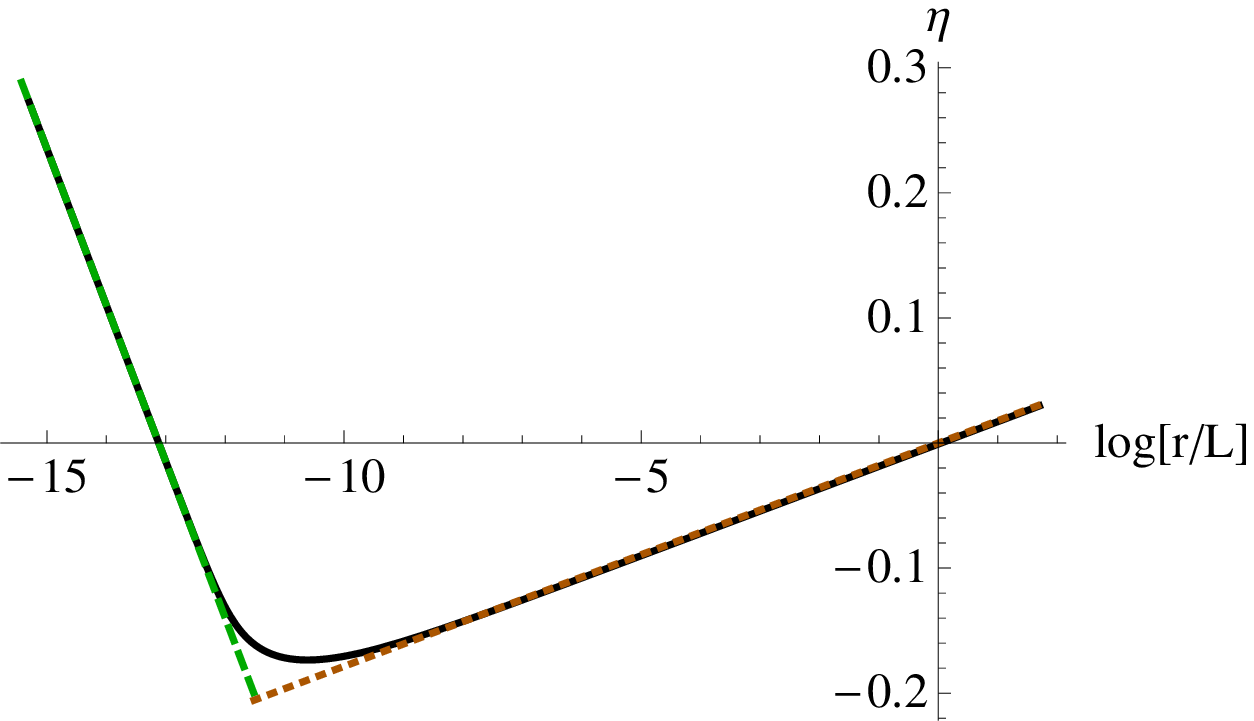}
\caption{\small Scalar $\eta$}
\end{subfigure}
~
\begin{subfigure}[b]{0.45\textwidth}
\includegraphics[width=\textwidth]{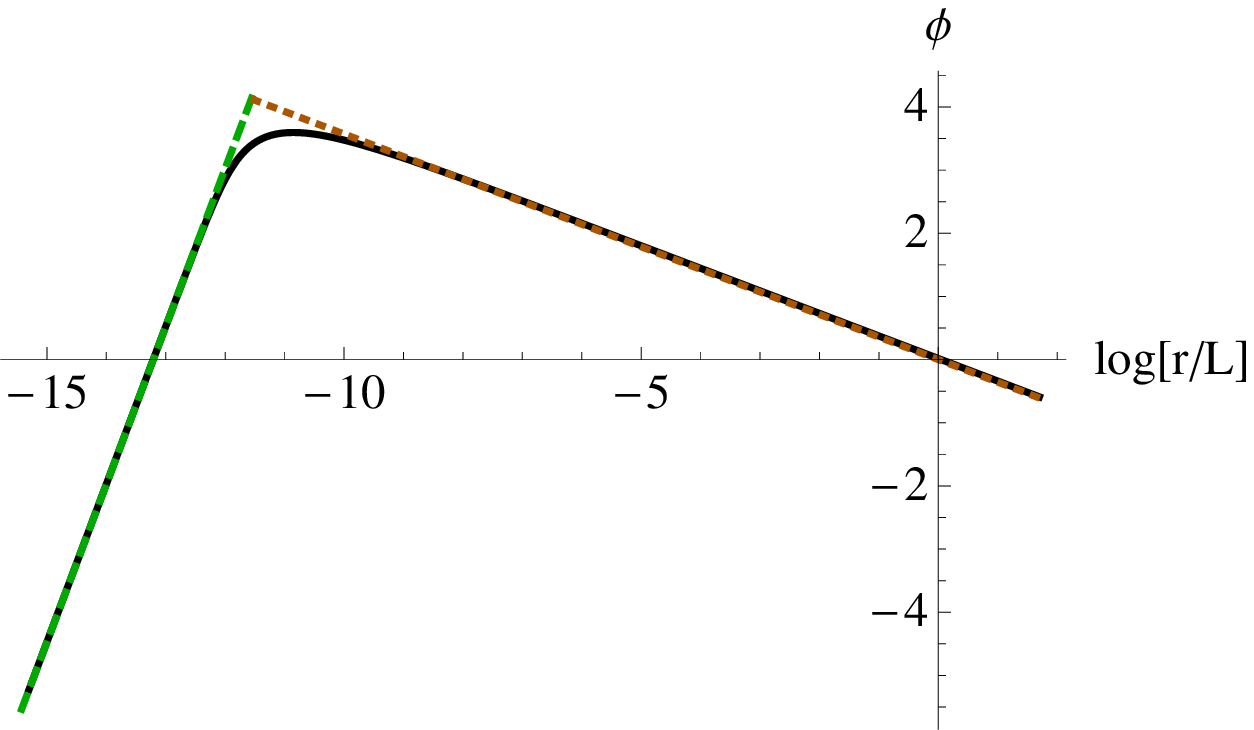}
\caption{\small Dilaton $\phi$}
\end{subfigure}\vspace{2mm}
\begin{subfigure}[b]{0.5\textwidth}
\includegraphics[width=\textwidth]{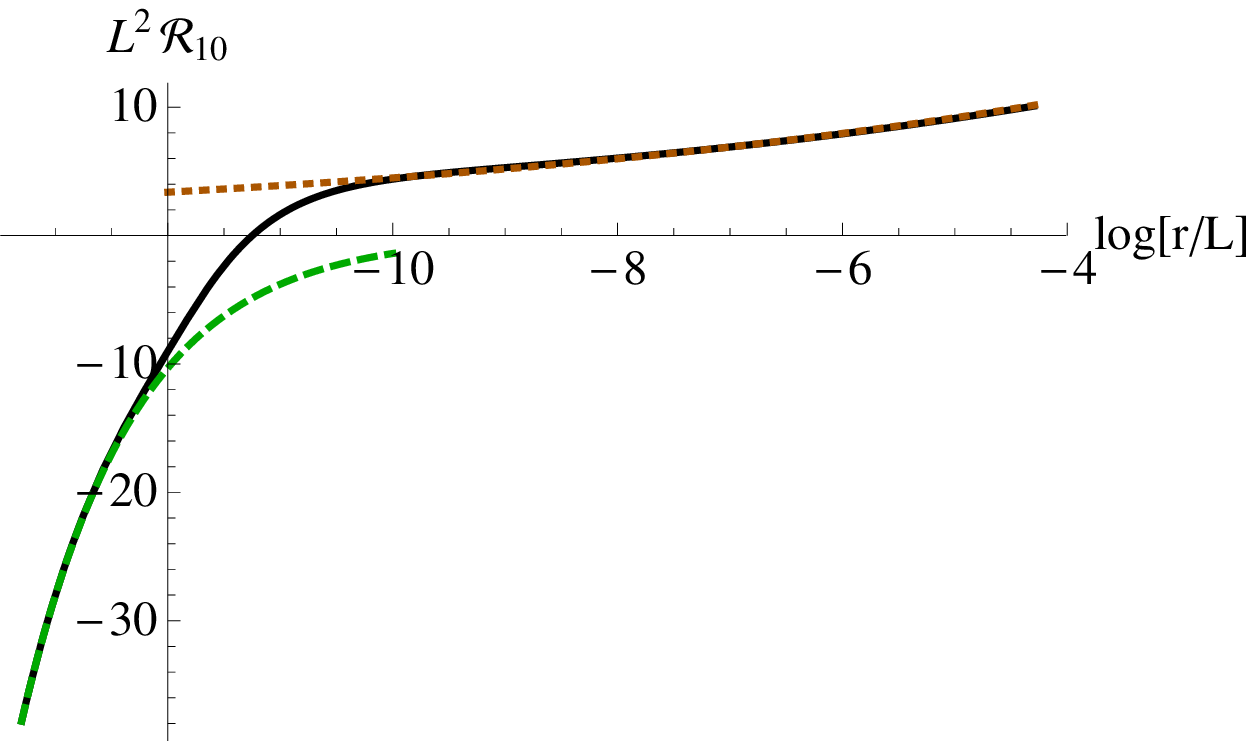}
\caption{\small 10D scalar curvature (string frame)}
\end{subfigure}
\caption{\small (Color online) Domain-wall solution for $d=3$ interpolating between the UV D2-brane solution \eqq{eq.Dpbranes} at $\log r/L \to \infty$ and the IR HV-Lif solution \eqq{IRgeneric} at $\log r/L \to -\infty$. The green, dashed lines on the left-hand side of the plots are  the  IR analytic expressions; the brown, dotted lines on the right-hand side are the UV values. } \label{fig.D2RG}
\end{center}
\end{figure}
The numerical result of the integration is given in Fig.~\ref{fig.D2RG}, where we provide the radial profile of the two scalars and the 10D  scalar curvature in string frame. In that figure we observe that the solution correctly interpolates between the UV asymptotic solution 
\eqq{eq.Dpbranes} and the IR solution \eqq{IRgeneric} with  $p=2$.

The different scales that characterize the RG flows were discussed in section \ref{sec.plessthan3}.
For densities ${\nq'}^{\mt{dual}} \ll \nq \ll \nq^\mt{dual}$ the ten-dimensional dilaton, or equivalently the size of the M-theory circle in Planck units, is large in the region $\uu \ll U \ll \ud$.
To verify that the eleven-dimensional supergravity description is valid we must also check that the eleven-dimensional curvature in Planck units is small.
Uplifting the solution \eqref{IRgeneric} (with $p=2$) to M-theory and computing the Ricci scalar in Planck units gives (after making use of \eqref{eq.relationIRradiusenergy})
\be
\ell_\mt{P}^2 \, {\cal R}_{11} \sim \frac{U^{7/3}}{\gym^2 \left(N^5\, \nq^2 \right)^{1/3}} \ ,
\ee
where the eleven-dimensional Planck length is given by $\lp = g_s^{1/3} \ls$. In the intermediate range under discussion this is indeed a small quantity (and the M-theory circle is large), and the 11D description is valid.

For small densities $\nq \ll {\nq'}^{\mt{dual}}$ one should match the uplifted solution to the localized M2 solution, which is beyond the scope of this paper.

\subsection{$d=2$ super Yang-Mills}

This case is very similar to the $d=3$ case. The main difference is that the starting point is Type IIB supergravity instead of Type IIA, and consequently the alternative description when the dilaton becomes large is obtained via S-duality instead of via an uplift to M-theory.

\begin{figure}[t]
\begin{center}
\begin{subfigure}[b]{0.45\textwidth}
\includegraphics[width=\textwidth]{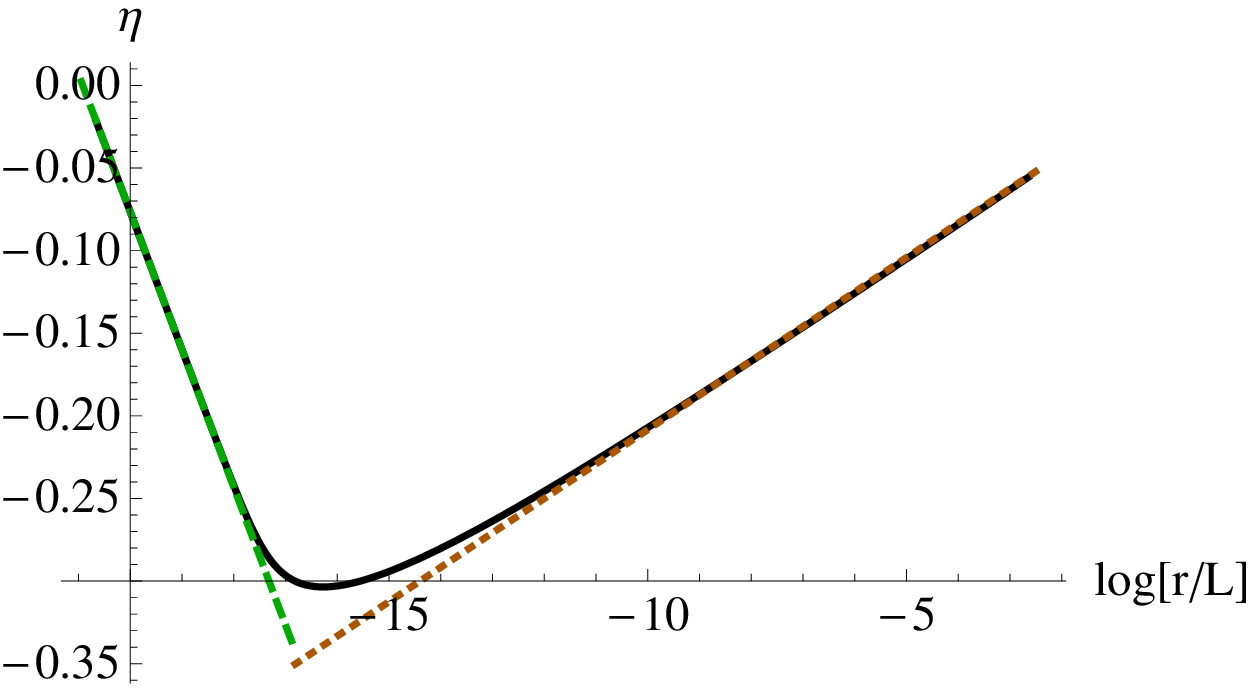}
\caption{\small Scalar $\eta$}
\end{subfigure}
~
\begin{subfigure}[b]{0.45\textwidth}
\includegraphics[width=\textwidth]{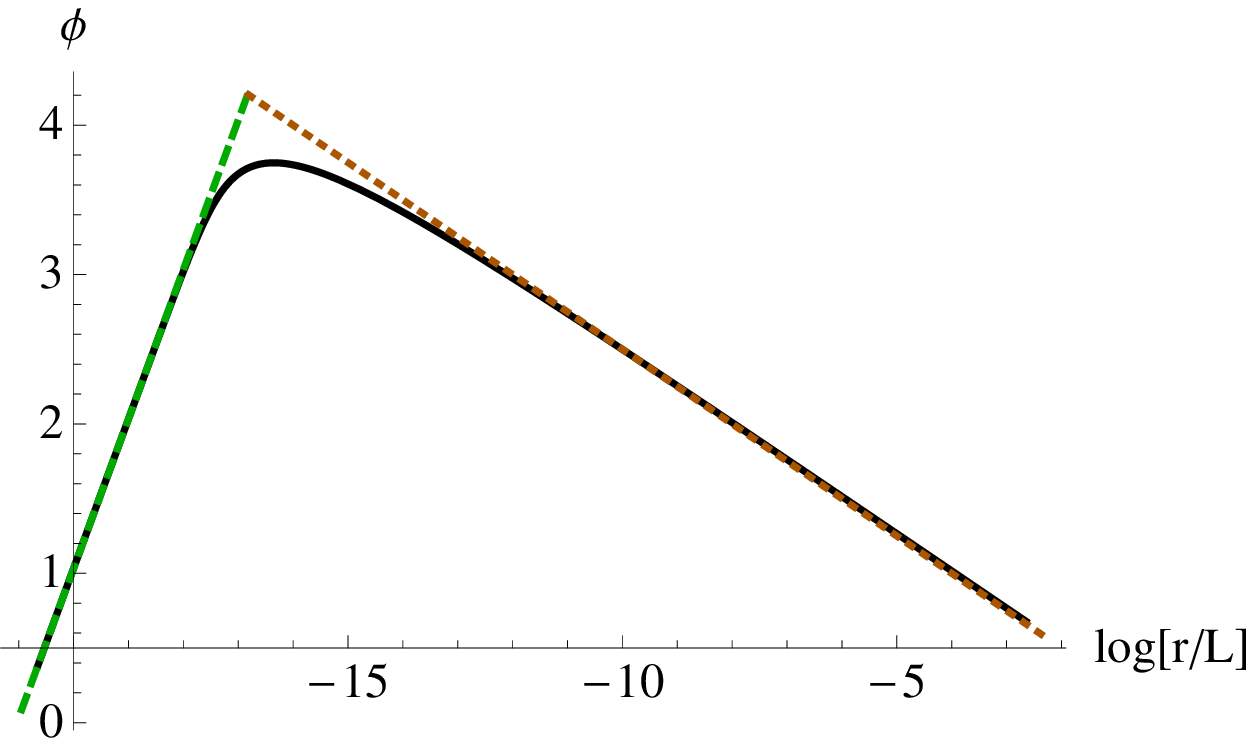}
\caption{\small Dilaton}
\end{subfigure}\vspace{2mm}
\begin{subfigure}[b]{0.5\textwidth}
\includegraphics[width=\textwidth]{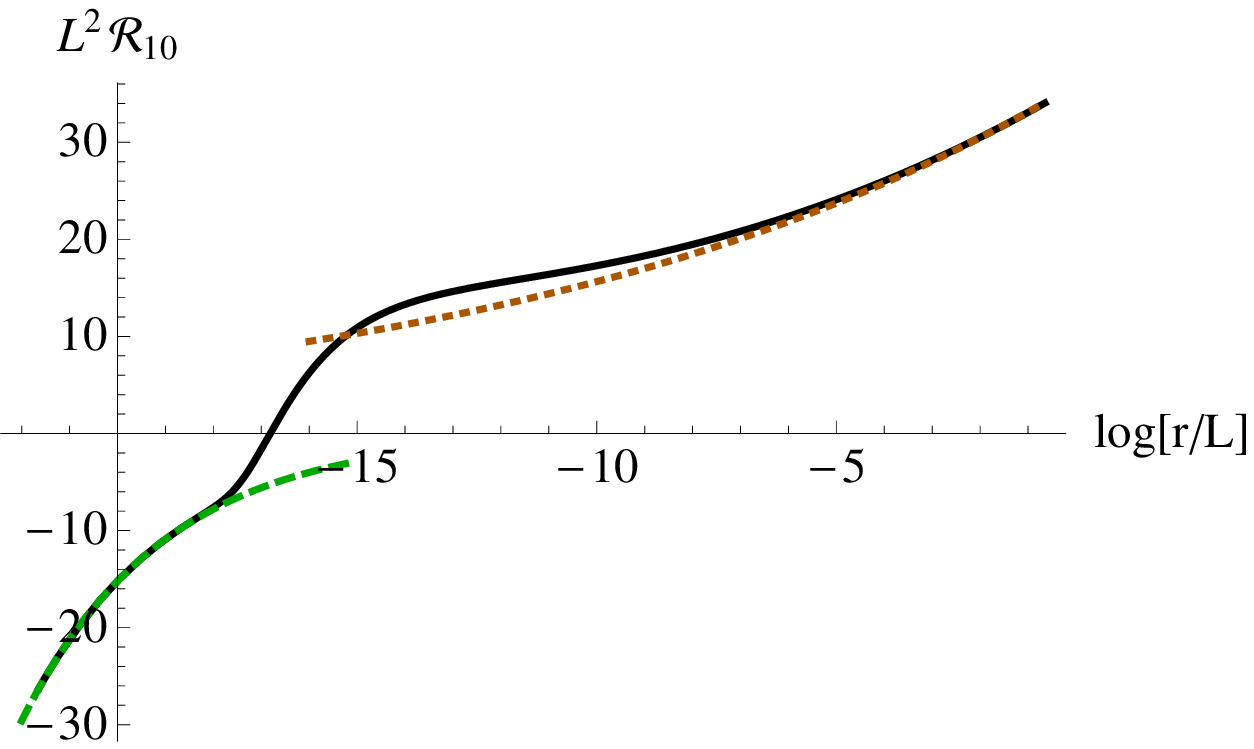}
\caption{\small 10D scalar curvature (string frame)}
\end{subfigure}
\caption{\small (Color online) Domain-wall solution for $d=2$. The color-coding is the same as in Fig.~\ref{fig.D2RG}.}\label{fig.D1RG}
\end{center}
\end{figure}
The modes that determine the domain wall solutions  are
\be
\Delta_\pm^1  = -\frac{7}{3} \pm \sqrt{\frac{7}{243} ( 2285 - 8 \sqrt{6679} ) } \ , \qquad
\Delta_\pm^2  = -\frac{7}{3} \pm \sqrt{\frac{7}{243} ( 2285 + 8 \sqrt{6679} ) } \ ,
\ee
and we provide the numeric integration in Fig.~\ref{fig.D1RG}.

The different scales that characterize the RG flows were discussed in detail in section \ref{sec.plessthan3}. For small densities, $ \nq \ll \nq^\mt{dual}$, the ten-dimensional dilaton becomes large in the region ${\uu \ll U \ll \ud}$ and an S-dual description is needed. Note that S-duality will act not just on the supergravity fields but also on the string sources.  One important feature in this respect, as we see from Eq.~\eqq{eq.Fpform} with $p=1$, is the presence in the initial solution of a space-dependent axion 
\be
C_0= - \frac{Q(x-x_0)}{L}  \,,
\ee
where $x_0$ is an integration constant. This has two important consequences. First, S-duality does not act simply  by reversing the sign of the dilaton. Second, the fundamental strings get transformed into D-strings that carry non-zero fundamental string charge. In order to account for this second feature, let us recall the action for (smeared) ($p,q$)-strings in Einstein frame (see for instance \cite{Cederwall:1997ts})
\begin{equation}
S_{(p,q)} = - \frac{N_q}{2\pi\ell_s^2} \int \left(\sqrt{q e^{-\phi}+e^{\phi}\left(p+q\, C_0\right)^2} \sqrt{-G_{tt}\,G_{rr}} \, \d t \wedge \d r - \left(p+q\,C_0\right)B_2+q\,C_2 \right) \wedge \Xi_8 \,.
\end{equation}
\noindent
For the values $p=1$, $q=0$ we recover the fundamental strings action we have been using. On the other hand selecting $p=0$, $q=1$ the action reduces to that of smeared D1-strings with non-zero fundamental string charge proportional to $C_0$. 

The backreacted solution in the S-dual frame is then found with the usual rules of S-duality. In particular, if $\tau=C_0+ie^{-\phi}$ is the axion-dilaton, performing the transformation $\tau\to-1/\tau$ we get the new values
\begin{equation}
e^{\phi}\,=\,\beta_\phi Q^{-1}\frac{r}{L}\left[\frac{1}{L^2}\left(x-x_0\right)^2+\beta_\phi^{-2}Q^4\left(\frac{L}{r}\right)^2\right]\,,\qquad C_0\,=\,\frac{\frac{1}{L}\left(x-x_0\right)}{\frac{Q}{L^2}\left(x-x_0\right)^2+\beta_\phi^{-2}Q^5\left(\frac{L}{r}\right)^2}\,.
\end{equation}
Under this transformation the Einstein-frame metric remains unchanged and the values of $B$ and $C_2$ are interchanged. We see that the S-dual solution allows us to extend somewhat the validity of the supergravity description in the vicinity of $x_0$, where the new dilaton is small. 

\subsection{$d=4$ super Yang-Mills}

\begin{figure}[t]
\begin{center}
\begin{subfigure}[b]{0.45\textwidth}
\includegraphics[width=\textwidth]{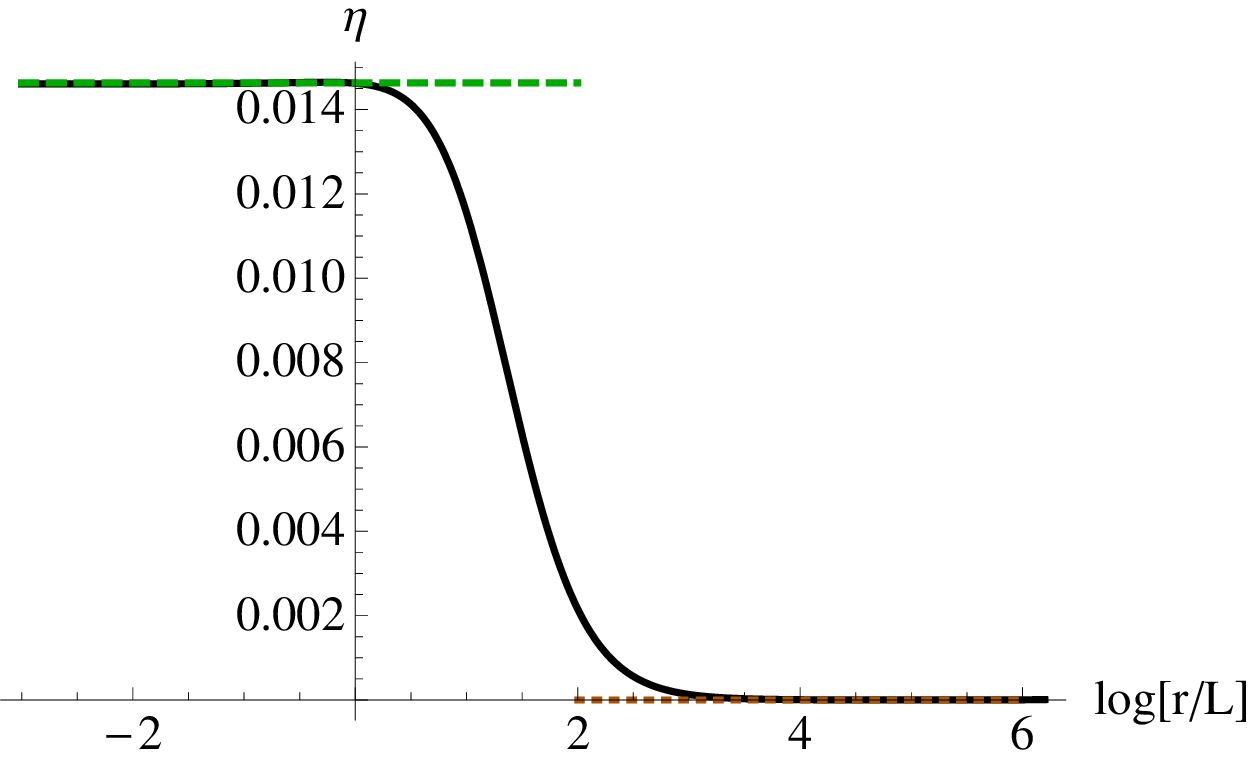}
\caption{\small Scalar $\eta$}
\end{subfigure}
~
\begin{subfigure}[b]{0.45\textwidth}
\includegraphics[width=\textwidth]{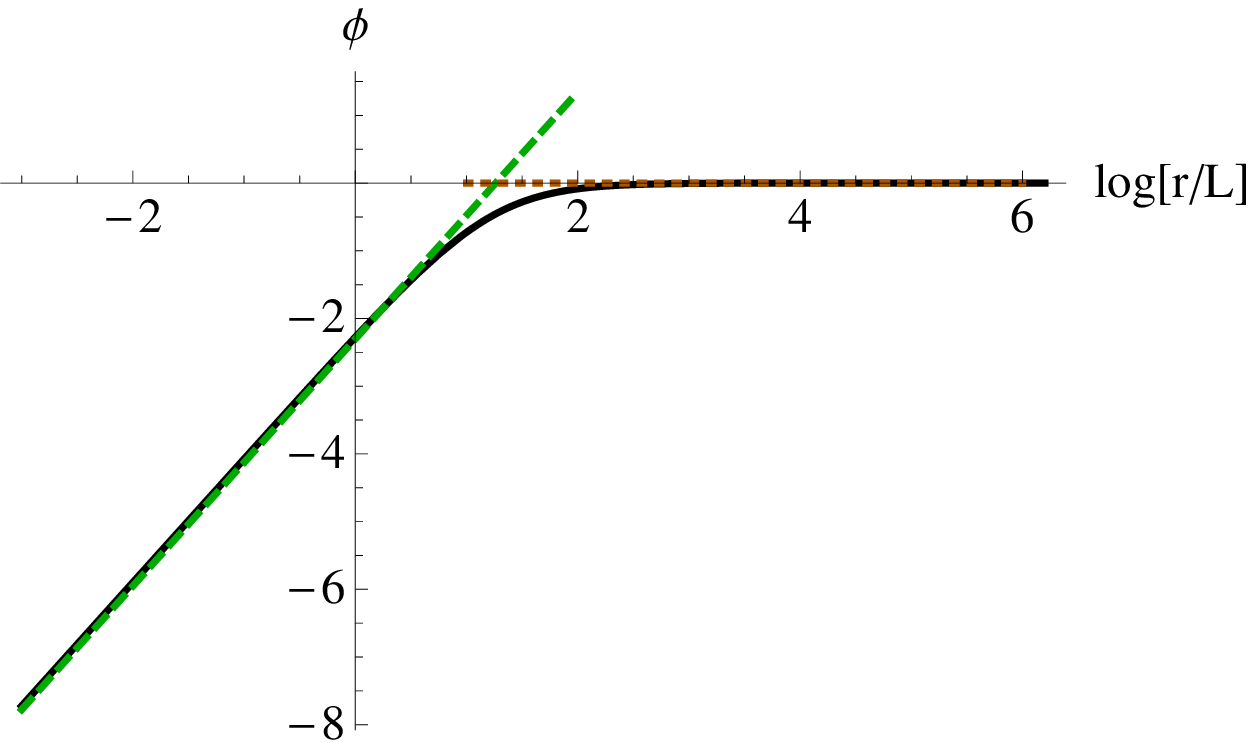}
\caption{\small Dilaton}
\end{subfigure}\vspace{2mm}
\begin{subfigure}[b]{0.5\textwidth}
\includegraphics[width=\textwidth]{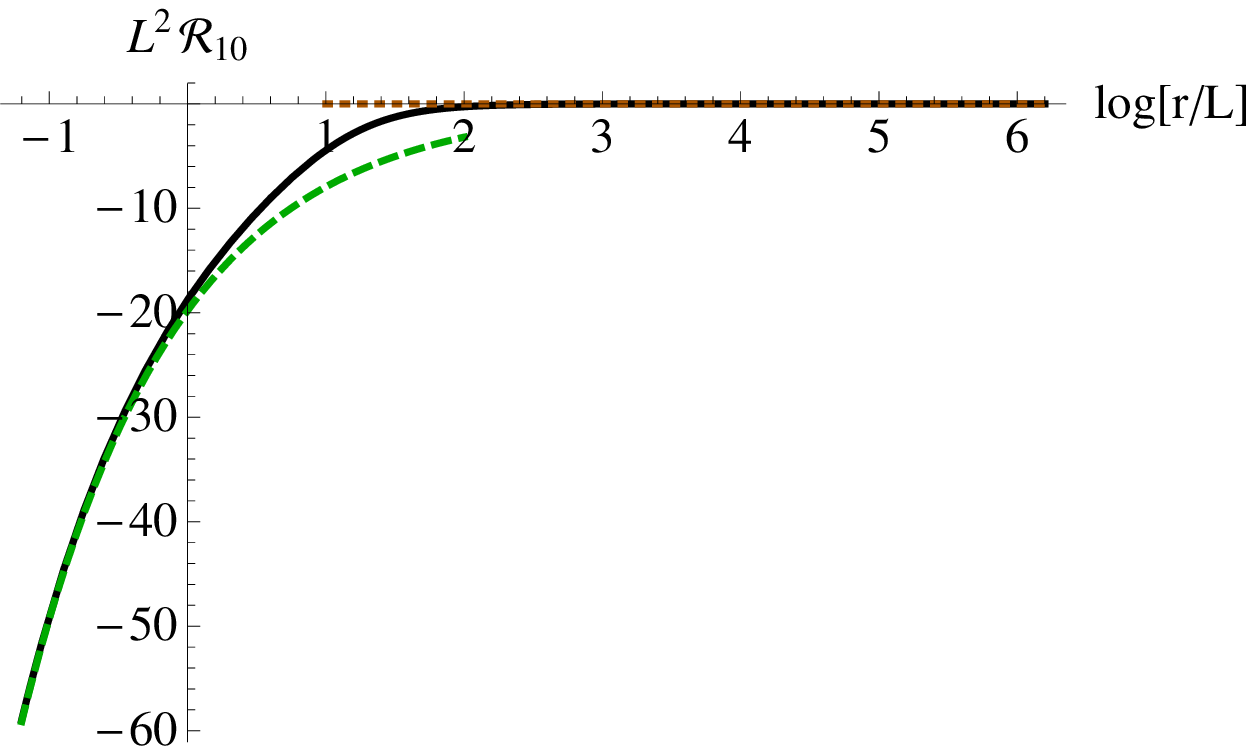}
\caption{\small 10D scalar curvature (string frame)}
\end{subfigure}
\caption{\small (Color online) Domain-wall for $d=4$ solution. The color-coding is the same as in Fig.~\ref{fig.D2RG}.}\label{fig.D3RG}
\end{center}
\end{figure}

The effects of a quark density in the RG flow of 3+1 SYM theory were studied originally in \cite{Kumar:2012ui}. We just comment on it for completeness.
The modes active in the IR of the domain wall are given by the powers
\be
\Delta_\pm^1  = - 5 \pm \sqrt{\frac{5}{17} ( 917 - 8 \sqrt{1279} ) } \ , \qquad
\Delta_\pm^2  = - 5 \pm \sqrt{\frac{5}{17} ( 917 + 8 \sqrt{1279} ) } \,.
\ee
Using these,  we build the numeric solution displayed in 
Fig.~\ref{fig.D3RG}.

In this case the theory in the absence of strings is conformal, therefore adding the quark density introduces the only scale in the problem.
The scales of the RG flow were discussed in section \ref{sec.pequals3}.

\subsection{$d=5$ super Yang-Mills}

\begin{figure}[t]
\begin{center}
\begin{subfigure}[b]{0.45\textwidth}
\includegraphics[width=\textwidth]{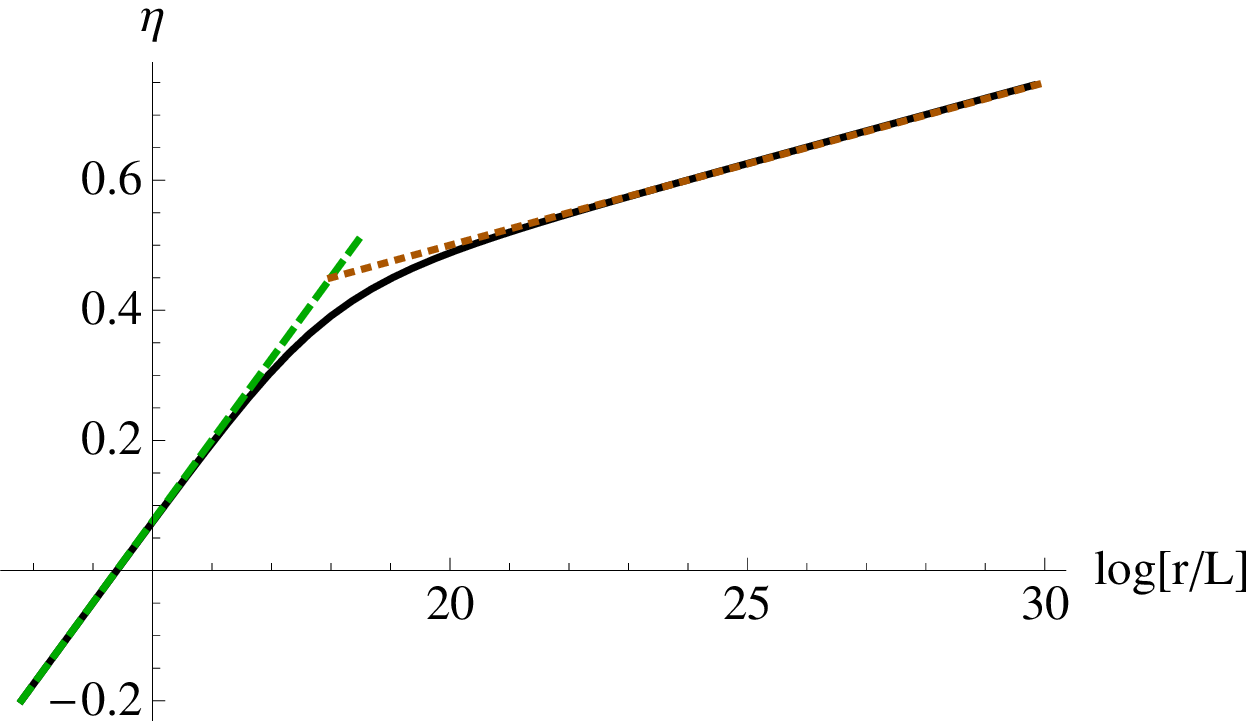}
\caption{\small Scalar $\eta$}
\end{subfigure}
~
\begin{subfigure}[b]{0.45\textwidth}
\includegraphics[width=\textwidth]{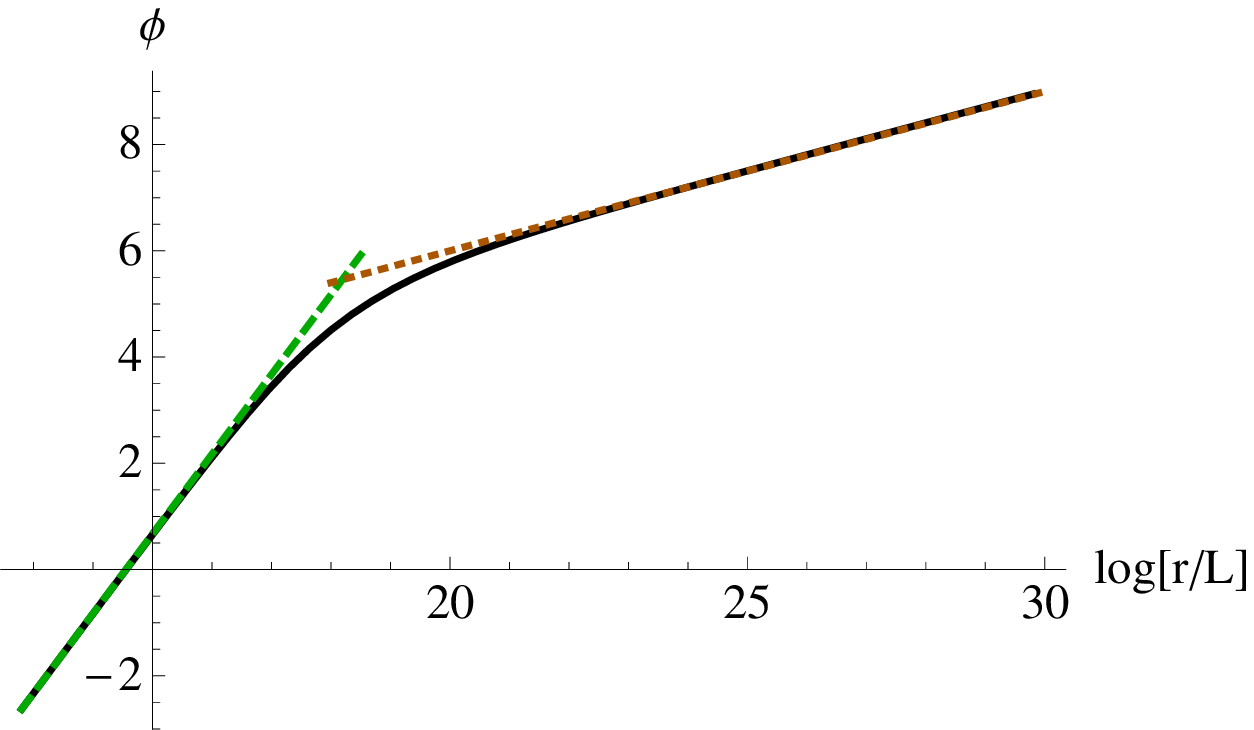}
\caption{\small Dilaton}
\end{subfigure}\vspace{2mm}
\begin{subfigure}[b]{0.5\textwidth}
\includegraphics[width=\textwidth]{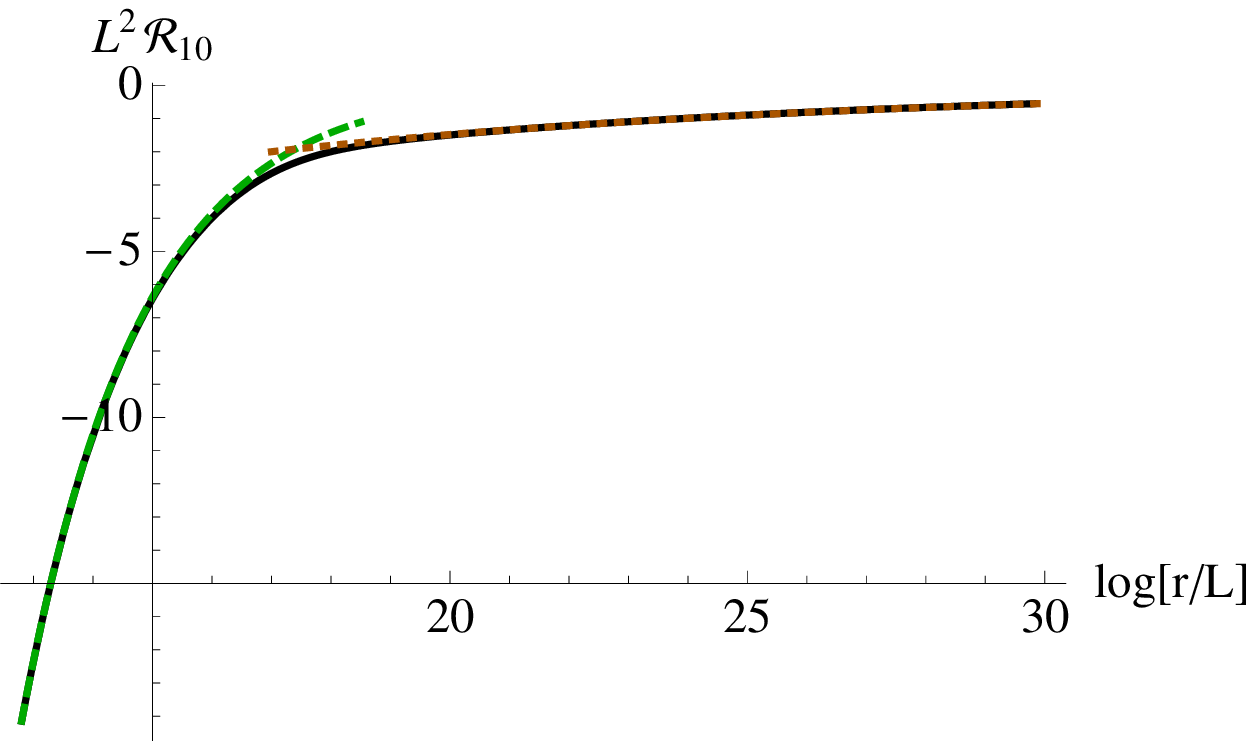}
\caption{\small 10D scalar curvature (string frame)}
\end{subfigure}
\caption{\small (Color online) Domain-wall solution for $d=5$. The color-coding is the same as in Fig.~\ref{fig.D2RG}.}\label{fig.D4RG}
\end{center}
\end{figure}

The modes that determine the domain wall solutions in this case are
\be\label{eq.D4Qmodes}
\Delta_\pm^1  = -1 \pm \sqrt{\frac{1}{3} (29 - 10)}\ , \qquad
\Delta_\pm^2  = -1 \pm \sqrt{\frac{1}{3} (29 + 10)} \ ,
\ee
and we provide the numeric integration  of the equations of motion in Fig.~\ref{fig.D4RG}.

The different scales that characterize the RG flow were discussed in section \ref{sec.pgreaterthan3}. 
For large densities $\nq\gg \nq^\mt{dual}$ there is a region of the flow in which the radius of the M-theory circle becomes large, and the IIA solution  does not provide the appropriate description. 
In this case one has to consider the M-theory uplift of \eqq{eq.pis4solution}-\eqq{IRscalars4}, which is given by the following   $AdS_3\times \mathbb{R}^4 \times S^4$ geometry:
\bse
\label{ads3}
\bal
\d s_{11}^2 &= -  \frac{r^2}{{\cal L}^2} \, f(r)\, \d t^2 + \frac{1}{{\cal Q}^2}\, \frac{r^2}{{\cal L}^2} \, \d \psi^2 +  \frac{{\cal L}^2}{r^2} \frac{\d r^2}{f(r)} + \d  x_4^2 + \frac{3}{2}\, {\cal L}^2 \, \d\Omega_4^2 \,, 
\label{eq.D4Qmetric} \\[2mm]
f(r) & = 1 - \frac{\rh^2}{r^2} \, , \\[2mm]
F_4 &= \frac{\sqrt{2}}{3^{1/4}} \,  \frac{1}{{\cal L}}\left[ \d x^1 \wedge \d x^2 \wedge \d x^3 \wedge \d x^4 \ +  \frac{3}{2}\, {\cal L}^4 \, \omega_4 \right] \,, 
\label{eq.D4Qform}
\end{align}
\ese
with $\psi$  the coordinate along the M-theory circle and
\be
{\cal L}^2 = \frac{2^{1/3}}{3^{1/2}}\, L^2 \ , \qquad {\cal Q}^2 = \frac{3^{5/2}}{2^{11/3}}  Q^2 \,.
\ee
The 11D Ricci scalar in Planck units is given by
\be
\label{constant}
\lp^2 \, {\cal R}_{11} = \frac{2 \lp^2}{{\cal L}^2} \sim \nc^{-2/3} \ ,
\ee 
and we conclude  that the eleven-dimensional description is reliable provided $\nc$ is large and $U \gg \uu$, which respectively ensure that, in Planck units, the curvature is small and the size of the eleven-dimensional circle is large. 

\begin{figure}[h!!!]
\begin{center}
\includegraphics[width=0.6\textwidth]{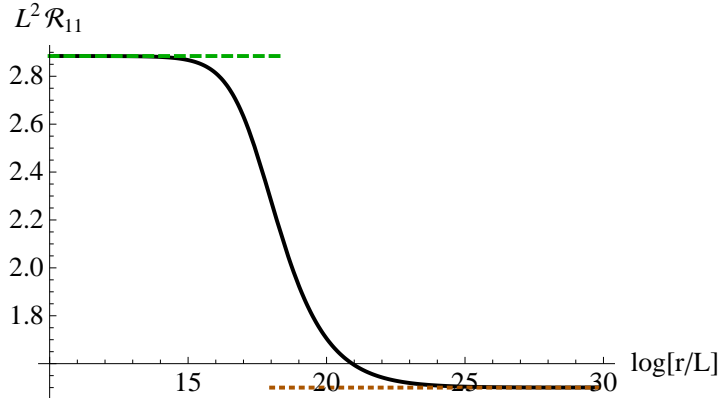}
\caption{\small (Color online) Domain-wall solution in M-theory interpolating between the UV $AdS_7 \times S^4$ solution at $\log r/L \to \infty$ and the IR solution $AdS_3\times \mathbb{R}^4 \times S^4$ at $\log r/L \to -\infty$.
} \label{fig.D4RGMtheory}
\end{center}
\end{figure}

Thus for supercritical flows with $\nq\gg \nq^\mt{dual}$ the solution is a domain-wall between the  $AdS_7\times S^4$ solution of M-theory in the UV, sourced by the M5-branes to which the D4-branes uplift, and the $AdS_3\times \mathbb{R}^4\times S^4$ solution.  We provide a plot of the corresponding eleven-dimensional Ricci scalar in  Fig.~\ref{fig.D4RGMtheory}. 
The 11D RG flow can therefore be interpreted as the backreaction on the M5-brane geometry of heavy M2-branes (to which the IIA strings uplift) oriented along the $\{ t,\psi, r \}$ directions of Eq.~\eqq{eq.D4Qmetric} and homogeneously distributed along the $\{x^1, \ldots, x^4\}$ directions. The presence of the M2-branes manifests itself in the $\d x^1 \wedge \ldots \wedge \d x^4$ term in $F_4$ in \eqq{eq.D4Qform}. This magnetic component is responsible for driving the flow between the asymptotically $AdS_7$ solution in the UV and the $AdS_3\times \mathbb{R}^4$ geometry in the IR. 

The reduction in the dimension of the $AdS$ factor along an RG flow triggered by a magnetic field is not  unfamiliar in the holographic context. For example, in \cite{D'Hoker:2009mm} a similar flow from $AdS_5$ to $AdS_3\times \mathbb{R}^2$ was found for ${\cal N}=4$ SYM theory in the presence of a magnetic field. In that case the fact that the IR is described by an effective (1+1)-dimensional CFT is due to the low-energy physics being governed by fermionic zero modes. It would be interesting to find a similar explanation in the case of the M5-brane theory. In our case, the modes \eqref{eq.D4Qmodes} correspond to sources and VEVs for two irrelevant operators in the IR, dual to scalars of masses ${m^2_1\, {\cal L}^2=2^{10/3}}$ and ${m^2_2\, {\cal L}^2=9\cdot 2^{4/3}}$ respectively. By choosing the positive roots, $\Delta^i_+$, we are switching on VEVs for those two operators.

\subsection{$d=6$ super Yang-Mills}

\begin{figure}[t]
\begin{center}
\begin{subfigure}[b]{0.45\textwidth}
\includegraphics[width=\textwidth]{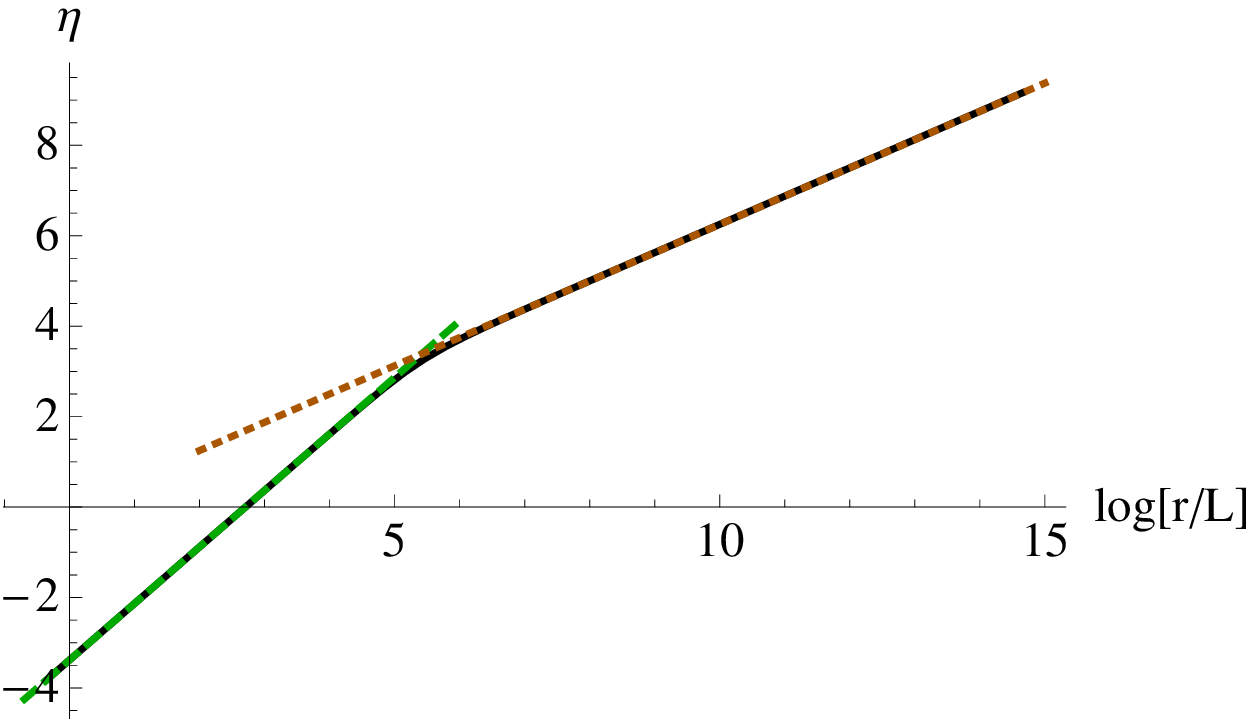}
\caption{\small Scalar $\eta$}
\end{subfigure}
~
\begin{subfigure}[b]{0.45\textwidth}
\includegraphics[width=\textwidth]{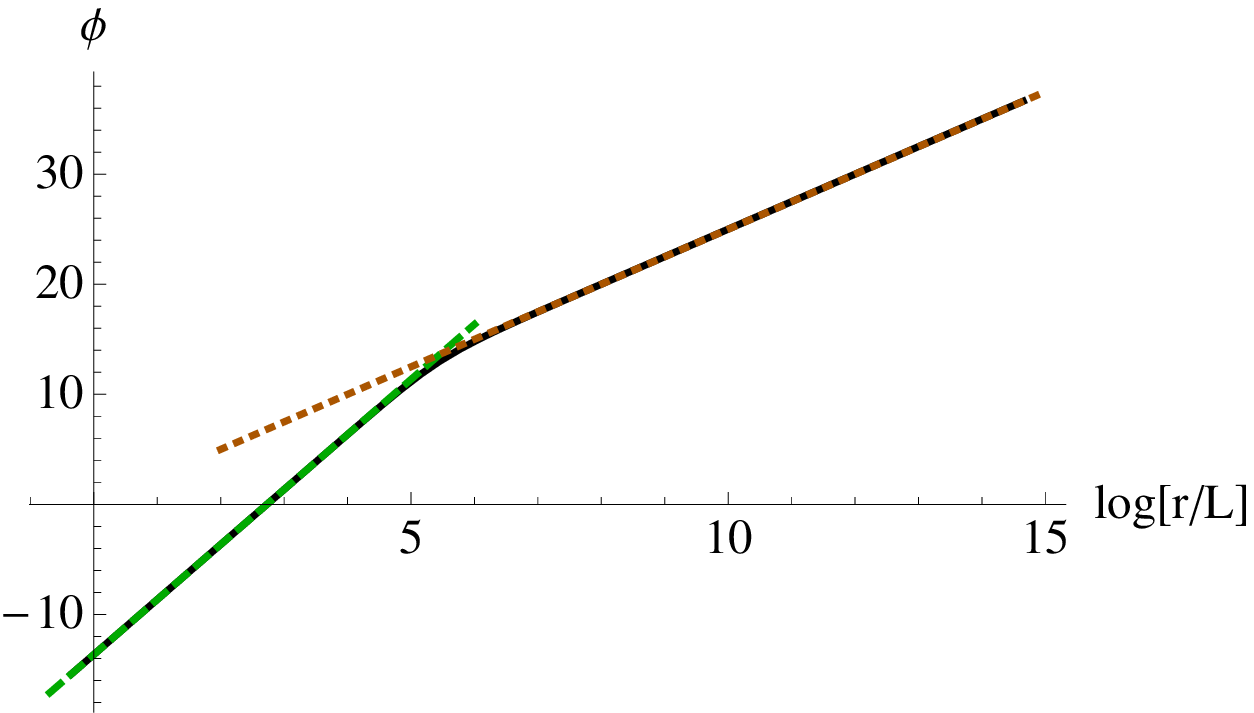}
\caption{\small Dilaton}
\end{subfigure}\vspace{2mm}
\begin{subfigure}[b]{0.5\textwidth}
\includegraphics[width=\textwidth]{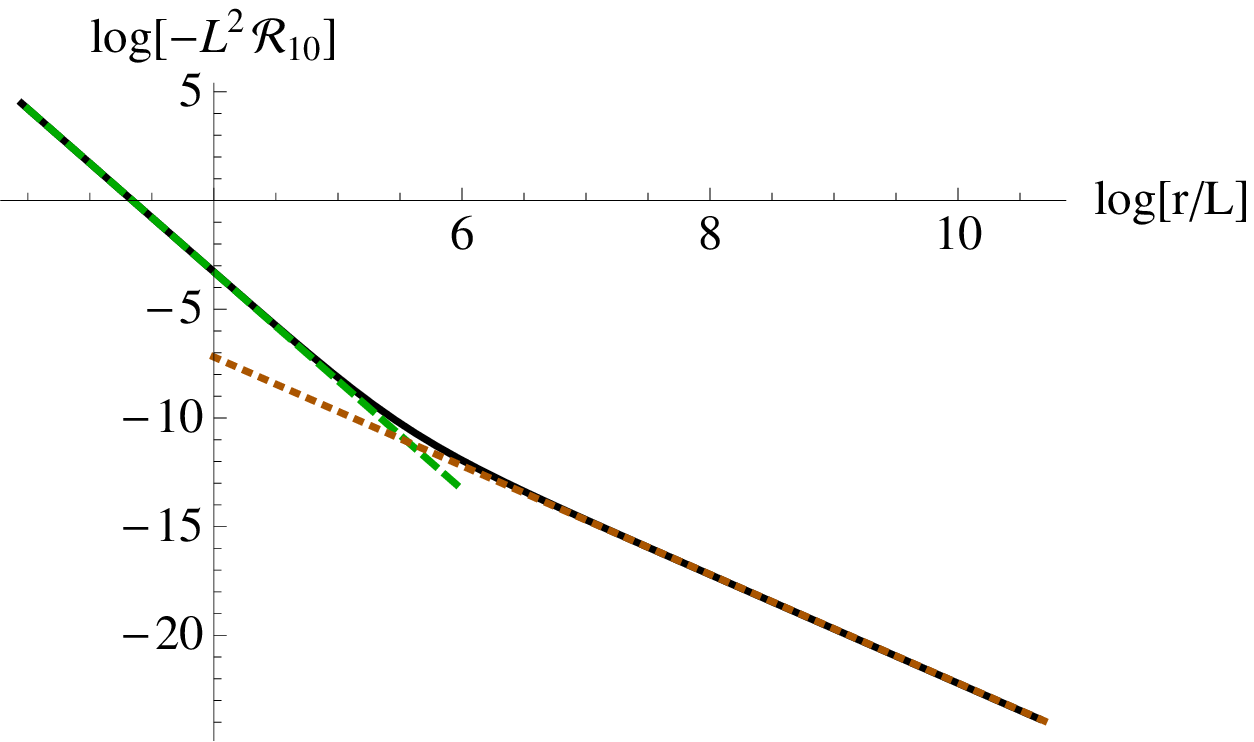}
\caption{\small Logarithm of the 10D scalar curvature (string frame)}
\end{subfigure}
\caption{\small (Color online) Domain-wall solution for $d=6$. The color-coding is the same as in Fig.~\ref{fig.D2RG}.}\label{fig.D5RG}
\end{center}
\end{figure}

For $p=5$, the relation \eqref{eq.relationIRradiusenergy} between the radius of the IR solution and the energy of the field theory implies that the IR corresponds to large values of $r/L$.
Therefore, in this section we invert the radial coordinate, taking $r\to L^2/r$. In the new radial coordinate, the modes that determine the domain wall in 5+1 SYM with quark density are
\be\label{eq.D5Qmodes}
\Delta_\pm^1  = - 3 \pm \sqrt{\frac{3}{7} ( 165 - 8 \sqrt{1279} ) }\ , \qquad
\Delta_\pm^2  = - 3 \pm \sqrt{\frac{3}{7} ( 165 + 8 \sqrt{1279} ) } \ ,
\ee
and we take the positive ones to integrate the equations numerically. The solution is displayed in Fig.~\ref{fig.D5RG}.

The several scales that characterize the RG flow are discussed in 
section \ref{sec.pgreaterthan3}.
For supercritical densities $\nq > \nq^\mt{dual}$ the crossover takes place in the region described by the S-dual configuration in terms of NS5-branes. Given that the original solution has vanishing axion, the dual configuration is easily found by taking $\phi\to-\phi$ and interchanging the NS and RR two-forms. This configuration with smeared D1 sources is thus supported by a non-vanishing $H$-field and a dilaton given by
\begin{equation}
e^{\phi}\,=\,\beta_\phi^{-1}\,Q\,\left(\frac{r}{L}\right)^{-5}\,.
\end{equation}

\section{Adding a heavy-quark density to other systems}\label{sec.other}
In this section we provide additional examples of systems with a gravity dual in which the addition of a quark density leads to the emergence of a scaling solution in the IR. In these cases we will present the IR solutions but  we will not settle the existence or otherwise of possible RG flows connecting them.

\subsection{D0-branes}\label{sec.D0branes}
The worldvolume theory associated to a stack of D0-branes in IIA supergravity is (0+1)-dimensional SYM, i.e.~a super quantum mechanical theory (SQM) with sixteen supercharges.  Since the dimensional reduction \eqref{eq.pplus2daction} is not well suited for the study of D0-branes, we provide in this section the analysis directly in ten dimensions.

The D0-brane metric and dilaton in the near-horizon limit in string frame are given by \eqq{UVmetric} and \eqq{UVdilatonDavid} with $p=0$, namely by
\be\label{eq.D0branes}
\d s^2 = \left(  \frac{u}{L} \right)^{-\frac{3}{2}} \left(-f(u)   \frac{u^{5}}{L^{5}} \d t^2 + \frac{L^2}{u^2} \frac{\d u^2}{f(u)} + L^2 \d \Omega_8^2 \right) \ , \qquad e^{\phi} = \left( \frac{u}{L} \right)^{-\frac{21}{4}} \ ,
\ee
with $f=1-(\uh/u)^7$.
This metric is conformal to $AdS_2\times S^8$, as can be seen via a change of radial coordinate.

Since in this case there are no spatial directions, the parameter $N_q$ labels the total number of quarks, as opposed to a quark density.  As follows from \eqq{eq.Fpform}, on the gravity side it induces a non-zero value for 
\be
F_0 = \frac{Q}{L} = \frac{2\kappa_{10}^2}{2\pi \ell_s^2} \, \frac{N_q}{7\,L^7\,V_8}   \ .
\ee
This means that we have turned on a Romans mass and that we are therefore working with massive Type IIA supergravity. With this value for the Romans mass, the equations of motion and Bianchi identities for the fluxes are satisfied.

As in previous sections, the introduction of quarks induces subleading corrections in the UV metric, and the UV asymptotic expansion  takes the form
\bse
\bal
\label{starts}
g_{uu} & = \left( \frac{u}{L} \right)^{-\frac{7}{2}} \ , \\
g_{tt} & = \left( \frac{u}{L} \right)^{\frac{7}{2}} \left[ 1 -  \frac{1253}{1053} \, Q \left( \frac{L}{u} \right)^6 + \varepsilon \left( \frac{L}{u} \right)^7 + {\cal O} \left( \frac{L}{u} \right)^{12} \right] \ , \\
g_{\Omega_8} & = \left( \frac{u}{L} \right)^{-\frac{3}{2}} L^2 \left[ 1 -  \frac{413}{2808} \, Q \left( \frac{L}{u} \right)^6 + \frac{1}{8}\, \varepsilon \left( \frac{L}{u} \right)^7 + {\cal O} \left( \frac{L}{u} \right)^{12} \right] \ , \\
e^\phi & = \left( \frac{u}{L} \right)^{-\frac{21}{4}} \left[ 1 -  \frac{385}{702} \, Q \left( \frac{L}{u} \right)^6 + \frac{7}{16} \,\varepsilon \left( \frac{L}{u} \right)^7 + {\cal O} \left( \frac{L}{u} \right)^{12} \right] \ .
\end{align}
\ese
This asymptotic expansion is similar to \eqref{eq.UVexpansion} in that the backreaction of the strings dominates over the term containing information about the one-point functions of the dual operators.
The main difference is that at order ${\cal O}(u^{-7})$ there is only one undetermined constant, $\varepsilon$, related to the energy of the dual theory.
The reason why there is no independent term corresponding to the normalizable mode of the dilaton, $v_\phi$, is that in 0+1 SYM theory the operator $\text{tr}\,F^2$ is proportional to the hamiltonian, and therefore $v_\phi \propto \varepsilon$.
At order ${\cal O}( u^{-14})$ there  is  another undetermined parameter, $v_{\eta}$, associated to the VEV of ${\cal O}_{14}$.

In the IR, there exists an exact solution whose metric and dilaton are given, in string frame, by
\be
\d s^2 = - f(r)   \frac{r^2}{L_\mt{IR}^{2}} \d t^2 + \frac{L_\mt{IR}^2}{r^2} \frac{\d r^2}{f(r)} + 28\, L_\mt{IR}^2 \, \d \Omega_8^2  \ , \qquad e^{\phi} = 2^6\, 7^3\, 5 \, \frac{L_\mt{IR}^7}{L^7} \ ,
\ee
where 
\be
f=\frac{(r-r_+)(r-r_-)}{r^2} \ , \qquad L_\mt{IR} = \left( \frac{9 }{2^{13} \, 7^{2} \, \pi^{4}}\, \frac{1}{Q} \right)^\frac{1}{8} L \ ,
\ee
and $r_+$ and $r_-$ are constants of integration. We see that this is solution is (black) $AdS_2\times S^8$ with a constant dilaton. 
In terms of the number of D0-branes and strings the radius of $AdS_2$ and the dilaton are given by 
\be
L_\mt{IR} = \left( \frac{9  }{2^{13} \, \pi^4} \, \frac{1}{\nq} \right)^{1/8}\,  \ell_s \sac 
g_s \, e^\phi = \left(  \frac{3^2 \, \, 5^{8/7} \, \, 7^{10/7}}{2^{43/7} \, \pi^4}  \, \frac{1}{\nq}  \right)^{7/8} \,\frac{1}{N} \ .
\ee
Note that there is no dependence on $\gym^2$ in these quantities, as expected from the fact that the system flows to a conformal phase. A quick calculation shows that the scalar curvature vanishes but that other curvature invariants scale as 
\be
\ell_s^4\, {\cal R}_{\mu\nu}{\cal R}^{\mu\nu} \sim \ell_s^4 \, {\cal R}_{\mu\nu\rho\sigma}{\cal R}^{\mu\nu\rho\sigma} \sim \nq^{1/2} \ .
\ee
For the massive IIA description to be valid we have to ensure that ${g_s\, e^\phi\ll1}$ and that ${\ell_s^4\, {\cal R}^2\ll 1}$. At large $\nc$ the second condition is more stringent than the first one and it requires that 
$N_q \ll \nc^2$. This condition is in tension with the condition that the backreaction of the strings in the UV be visible in the classical supergravity approximation, which requires that $N_q={\cal O}(\nc^2)$. This suggests that the IR solution that we have constructed might not be the endpoint of the RG flow that starts from the UV solution \eqq{starts}. Although we have not settled this issue completely, preliminary numerical investigations seem to confirm this possibility.

\subsection{Brane intersections}

The scaling solutions in the IR of the D$p$-branes can be generalized to the case of D$p$/D$q$-intersections in the presence of a density of fundamental strings. We will consider D$p$-branes delocalized in the worldvolume of the D$q$-branes. Moreover, starting from the D$p$/D$(p+4)$ intersection, other cases can be obtained by T-duality along one of the coordinates transverse to the D$p$ but parallel to the D$(p+4)$. In this way one gets D$(p+1)$/D$(p+3)$ and, iterating, D$(p+2)$/D$(p+2)$ intersections. Since the sources do not break the isometry employed to dualize, the properties of the solution will be preserved, in particular the Lifshitz and hyperscaling-violating coefficients. 

It is thus sufficient to consider the following setup
   \bal
\begin{array}{ccccccccc}
&x^1&\cdots&x^p&x^{p+1}&\cdots&x^{p+4}&r&{S}^{4-p}\\
{\rm D}p&\times&\times&\times&-&-&-&-&-\\[2mm]
{\rm D}(p+4)&\times&\times&\times&\times&\times&\times&-&-\\[2mm]
{\rm F}1&-&-&-&-&-&-&\times&-\nonumber
\end{array}
\end{align}
The properties of the solution will solely depend on the dimension of the intersection, that is, the value of $p$. The geometry is supported generically by the forms
\be
F_{8-p}=\left(3-p\right)L^{3-p}\,\d x^{p+1} \wedge \cdots \wedge \d x^{p+4}\wedge\omega_{4-p}\,,\qquad\quad\quad F_{4-p}=\left(3-p\right)L^{3-p}\,\omega_{4-p}
\ee
where the quantization condition gives in this case $L^{3-p}\sim N$, with $N$ the number of $(p+4)$-branes. We have also rescaled $x^p,\dots,x^{p+4}$ to absorb an additional parameter in $F_{8-p}$ and express everything just in terms of $L$.

The fundamental strings are distributed according to the smearing form
\be
\Xi_8 = \frac{1}{V_{4-p}} \d x^1 \wedge \cdots \wedge \d x^{p+4} \wedge \omega_{4-p} \ .
\ee

This activates further components in the forms 
\be\label{pqflux}
F_p=\left(-1\right)^{\left[\frac{p+1}{2}\right]}\,\frac{Q}{L}\,\d x^{1} \wedge \cdots \wedge \d x^{p}\,,\qquad\qquad F_{p+4}=\left(-1\right)^{\left[\frac{p+1}{2}\right]}\,\frac{Q}{L}\,\d x^{1} \wedge \cdots \wedge \d x^{p+4}\,,
\ee
where the dimensionless parameter $Q$ is again related to the spatial density of strings as
\be
Q\,=\,\frac{2\kappa_{10}^2}{2\pi\ell_s^2}\,\frac{L^{p-2}}{(3-p)V_{4-p}}\, N_q \ .
\ee
Now the strings have two options to form baryons: either a (4-$p$)-brane wrapping $S^{4-p}$ or alternatively a (8-$p$)-brane along $\mathbb{R}^4\times S^{4-p}$, reflected in the choice of flux (\ref{pqflux}).  A combination of both types of baryons is also admissible. 

It is possible to compactify the ten-dimensional action with the mentioned fluxes on $S^{4-p}$. If one subsequently reduces on $\mathbb{R}^4$ (or more properly its compact version), one gets a $(p+2)$-dimensional action with a modified potential. For $Q=N_q=0$ this action admits the usual D$p$/D$(p+4)$ intersection solution. When $Q\ne0$ and $p=0, 1$ there is as well an exact scaling solution of the form (\ref{IRgeneric}). There are no exact hyperscaling geometries for $p\ge2$. This may be related to the remark that $p+4\ge6$, and as we observed there are no solutions of that type for D$p$-branes alone with such high $p$. 

The IR solution for $p=1$ reads, in Einstein frame
\be\label{D1D5metric}
\d s^2\,=\,\left(\frac{r}{L_\mt{IR}}\right)^\frac12\left[-\frac{r^4}{L_\mt{IR}^4}f(r)\,\d t^2+\frac{r^2}{L_\mt{IR}^2}\d x_1^2+\frac{L^2_\mt{IR}}{r^2} \frac{\d r^2}{f(r)}+\beta_x^2\,Q^\frac25\,\frac{L_\mt{IR}}{r}\d x_4^2+\frac{L_\mt{IR}^2}{3}\d \Omega_3^2\right]\,,
\ee
where the blackening factor, the IR radius and the dilaton are
\be
f\,=\,1-\frac{\rh^3}{r^3}\,,\quad\quad\quad L_\mt{IR}\,=\,\left(\frac{\beta_L}{Q}\right)^{\frac15}\,L\,,\quad\quad\quad e^{\phi}\,=\,\beta_\phi\,\left(\frac{L_\mt{IR}}{L}\right)^4\,\frac{r}{L_\mt{IR}}\,.
\ee
The unspecified numerical factors $\beta_X\,=\,\left\{\beta_x, \beta_L, \beta_\phi\right\}$ depend on the election of flux in (\ref{pqflux}). For the solution supported by $F_1$ one has 
\be
\beta_X\,=\,\left\{\frac{3^{3/40} \, 7^{9/40}}{2^{1/10} \, 5^{1/4}}, \frac{2^3 \, 3^{3/2}}{7^{1/2}}, \frac{1}{12}\right\} \,,
\ee
while in case the five-form is switched on the numbers read 
\be
\beta_X\,=\,\left\{\frac{ 7^{1/8}}{2^{1/10} \, 3^{1/40} \, 5^{1/20}}, \frac{72}{5}, \frac{25}{252}\right\}\,.
\ee
The metric (\ref{D1D5metric}), once reduced to 
the three-dimensional Einstein frame, is a Lifshitz space with $z=2$ and no hyperscaling violation.

Interestingly, the solution for $p=0$, identified with the D$0$/D$4$ intersection at finite density, is $AdS_2$ with constant dilaton. As explained, this solution is supported by a four-form on the four-sphere, an eight-form on $\mathbb{R}^4\times S^{4}$ and additionally either $F_0$ or extra components of $F_4$ along $\mathbb{R}^4$. In this second instance it is possible to uplift the solution to M-theory. Due to the non-vanishing two-form (dual to $F_8$) the metric in eleven dimensions is a $U(1)$ fiber over $AdS_2$, which is to say, AdS$_3$ in global coordinates:
\be
\d s^2\,=\,\frac{\mathcal{L}^2}{4}\left[-\cosh^2{\sigma}\,\d t^2+\d\sigma^2+\left(\frac{1}{\mathcal{L}\mathcal{Q}^{1/2}}\d\psi+\sinh{\sigma}\d t\right)^2+6\,\d\Omega_4^2\right]+\frac{2^{3/2}}{3^{1/4}}\mathcal{Q}^{1/2}\,\d x_4^2\,.
\ee
The new radius and charge are related to the ones defined above as
\be
\mathcal{L}\,=\,\frac{L}{3^{1/6}}\,,\qquad\qquad\qquad\mathcal{Q}\,=\,\frac{3^{5/6}}{2^{4}}\,Q\,.
\ee
This solution is sourced by a distribution of M$2$-branes that span the $AdS$ directions. In terms of the number of M$5$-branes, $N$, inherited from the uplift of the D$4$-branes, the radius goes as $\mathcal{L}\sim N^{\frac13}$. The dependence of the curvature in this parameter is ${\cal R}=\frac{2}{\mathcal{L}^2}\sim N^{-\frac23}$ and is accordingly small at large $N$.

\subsection{ABJM}\label{sec.ABJMstrings}

So far we have explored introducing a density of heavy-quarks to a large class of field theories arising from various brane configurations by sourcing supergravity with a smeared distribution of fundamental strings. The dual gauge theories are typically large-$N$ (super) Yang-Mills theories. There is another class of large-$N$ gauge theories which are Chern-Simons theories with matter fields. The prototypical example is the so called ABJM model \cite{Aharony:2008ug}, which is an $SU(N)_k \times SU(N)_{-k}$ (superconformal) Chern-Simons theory with matter, where $k$ denotes the Chern-Simons level. This is the worldvolume theory of $N$ coincident M2-branes probing a ${\mathbb C}^4/{\mathbb Z}_k$ singularity.

In the large $N$-limit, the gravity dual is given by the $AdS_4 \times S^7/{\mathbb Z}_k$ background in eleven-dimensional supergravity. One can represent the $S^7/{\mathbb Z}_k$ manifold as an $U(1)$ fibration over the K\"{a}hler-Einstein base ${\mathbb {CP}}^3$. Then reducing on the fiber, we arrive to a Type IIA description of $AdS_4 \times {\mathbb {CP}}^3$ which is valid at large $N$ and $k$, maintaining the ratio $N/k$ fixed within the window $k\ll N\ll k^5$. The corresponding ten-dimensional geometry and fluxes, in the string frame, are given by
\bse\label{abjm}
\bal 
 \d s^2 &=  - \left(\frac{u}{L_\mt{UV}}\right)^2 \d t^2 + \left(\frac{u}{L_\mt{UV}}\right)^2 \d x_2^2 +  \left(\frac{L_\mt{UV}}{u}\right)^2 \d u^2 + \frac52 \,L_\mt{UV}^2 \, \d s_{{\mathbb {CP}}^3}^2 \ , \label{abjm1} \\[2mm]
F_6 &= 5\,L^5 \, \label{abjm2}\omega_6 \ ,\\[2mm]
F_2 &= \frac{4}{15}\,k_\mt{UV}\, L\,J \ , \\[2mm]
e^\phi &=\frac{75}{8}\,{k_\mt{UV}^{-5/4}} \label{abjm4}
\end{align}
\ese
where $J$ is the K\"{a}hler form on ${\mathbb {CP}}^3$, normalized as $J^3= 6\,\omega_6$. This is a somewhat unusual way of writing this solution, picked for consistency with the rest of the paper and also to make explicit the dimensionality of the different quantities. First of all, the metric of the internal manifold is normalized as in the previous sections. Moreover, as it stands, (\ref{abjm2}) coincides with (\ref{eq.F8minusp}), which ensures that $L$ is related to the number of D2-branes $N$ as in (\ref{quantcond}). The parameter in the two-form flux is not the quantized one $k$, but connected to it as $k_\mt{UV}\sim k\,N^{-1/5}$, the proportionality constant being unimportant for our purposes. Finally, in our conventions the radius appearing in the metric is related to the one in the fluxes by  $L_\mt{UV}^4=L^4\,k_\mt{UV}^{-1}\sim N\,k^{-1} $.

In previous sections we studied a scaling solution of Type IIA supergravity with a quark density in which the internal space is $S^6$. An analogous solution exists in the case in which the internal space is 
${\mathbb {CP}}^3$. This can be recovered from our general discussion, but we write it here, in string frame, for completeness:
\bse \label{eqnspl}
\bal
\d s^2 &= \left(\frac{r}{L}\right) \left[- \left(\frac{r}{L}\right)^{10} \d t^2 + \left(\frac{r}{L}\right)^2 \d x_2^2 +\beta_\ell^2\,Q^{-1/3}\left(\frac{L}{r}\right)^2 \d r^2 + \beta_\eta\,Q^{-1/3}\,L^2 \, \d s_{{\mathbb {CP}}^3}^2 \right] , \label{eqnspl1}  \\[4mm]
e^\phi&=\beta_\phi\,Q^{-5/2}\,\left(\frac{r}{L}\right)^{5/2}\,.
\end{align}
\ese
The fluxes take the form stated in (\ref{eq.F8minusp}) and (\ref{eq.Fpform}) with $p=2$. The numerical factors are 
\be
\left\{\beta_\ell, \beta_\phi, \beta_\eta\right\}=\left\{\frac{2^{4/3}\, 3^{1/2}}{5^{1/3}\, 7^{1/6}},\frac{2^{5/3}\, 5^{1/3}\, 7^{1/6}}{11^{1/2}}, \frac{2^{2/3}\, 5^{1/3}}{7^{1/3}}\right\} \,.
\ee
This corresponds, in four-dimensional Einstein frame, to the expected scaling geometry with $z=5$ and $\theta=1$ described in (\ref{eq.genericscalinginit}).

Evidently, the major difference between the solutions \eqref{abjm} and \eqref{eqnspl} is that in the former we have $k_\mt{UV} \not = 0$ and $N_q = 0$, whereas in the latter we have $k_\mt{UV} = 0$ and $N_q \not = 0$. We can ask what happens when both $N_q \not = 0$ and $k_\mt{UV} \not = 0$, i.e.~when we add a quark density to the ABJM solution. In this case, the equations of motion do not admit an exact scaling solution. Nevertheless, we can check the UV and the IR asymptotics. To that end, we take a general ansatz that comprises both geometries. We need a two form that includes both contributions, 
\be
F_2 =  \frac{4}{15}\,k_\mt{UV}\, L\,J - \frac{Q}{L} 
\,\d x^1 \wedge \d x^2 \,,
\ee
together with metric components and a dilaton with general radial dependence. The six-form coincides for both backgrounds.

It can be seen that the ABJM UV asymptotics is preserved at finite density.  The various functions in the solution take the following asymptotic form:
\bse
\label{abjmUV}
\bal
g_{uu}& = \left(\frac{L_\mt{UV}}{u}\right)^2 \ , \\[2mm] 
g_{tt} & =- \left(\frac{u}{L_\mt{UV}}\right)^2 \left[ 1 -  \alpha_t \,Q\,\left(\frac{L_\mt{UV}}{u}\right)^2-\varepsilon\left(\frac{L_\mt{UV}}{u}\right)^3 +\mathcal{O}\left(\frac{L_\mt{UV}}{u}\right)^4 \right]\ , \\[2mm]
g_{xx} &=\left(\frac{u}{L_\mt{UV}}\right)^2 \left[ 1 + \alpha_x \,Q\,  \left(\frac{L_\mt{UV}}{u}\right)^2+ \frac{\varepsilon}{2} \left(\frac{L_\mt{UV}}{u}\right)^3  +\mathcal{O}\left(\frac{L_\mt{UV}}{u}\right)^4 \right] \ , \\[2mm]
g_{{{\mathbb {CP}}^3}}& = \frac{5}{2} L_\mt{UV}^2  \left[ 1 +  \alpha_c \,Q\,  \left(\frac{L_\mt{UV}}{u}\right)^2 +\frac{4\,v_\phi}{9}\,\left(\frac{L_\mt{UV}}{u}\right)^4 + v_\eta\,\left(\frac{L_\mt{UV}}{u}\right)^6+\mathcal{O}\left(\frac{L_\mt{UV}}{u}\right)^7  \right] \ , \\
 e^{\phi}& =\frac{75}{8}\,{k_\mt{UV}^{-5/4}}\left[1 + \alpha_\phi \,Q\,  \left(\frac{L_\mt{UV}}{u}\right)^2 +v_\phi\,\left(\frac{L_\mt{UV}}{u}\right)^4 + \frac{v_\eta}{2}\,\left(\frac{L_\mt{UV}}{u}\right)^6+\mathcal{O}\left(\frac{L_\mt{UV}}{u}\right)^7 \right] \,. 
\end{align}
\ese
Here $\alpha_t$, $\alpha_x$, $\alpha_c$ and $\alpha_\phi$ are four constants which can be determined from the linearized equations of motion and $g_{{\mathbb {CP}}^3}$ is the overall metric component in front of the internal directions.\footnote{This corresponds roughly to $e^{2\eta}$ used in the previous sections, but one has to keep in mind the change of frame.} The free parameters $\varepsilon$, $v_\phi$ and $v_\eta$ correspond to the VEVs of the stress-energy tensor and the operators dual to the dilaton and the volume of the internal manifold, respectively. 

It is clear from the  expansion \eqq{abjmUV} that the quark density induces a subleading correction in the UV to the ABJM solution \eqq{abjm}. Conversely, we will now show that the `$k_\mt{UV}$-component' of the flux induces a subleading  correction to the IR solution \eqq{eqnspl}. Since the two-form flux enters the equations of motion quadratically, the correction caused by it will enter at order $k_\mt{UV}^2$.\footnote{In the series expansion, an integer power of $k_\mt{UV}$ will be multiplied by an appropriate power of the radius and this product is the physical expansion parameter for the system.} Thus, the linearized solution will now take the form:
\bse
\bal
& g_{rr}(r) = \frac{L}{r} \, \beta_\ell^2\, Q^{-1/3} \ , \\
& g_{tt}(r) =- \frac{r^{11}}{L^{11}} \left( 1 +  \delta  g_{tt}(r) \right) \ , \\
& g_{xx} (r) = \frac{r^{3}}{L^3} \left( 1 + \delta  g_{xx}(r) \right) \ , \\
& g_{{\mathbb {CP}}^3} (r) = \beta_\eta\, \frac{r}{L} \, Q^{-1/3} \left( 1 + \delta g_{{\mathbb {CP}}^3}(r) \right) \ , \\
& \phi = \frac{5}{2} \log\left[ \frac{r}{L} \right] +  \log\left[ \beta_\phi\, Q^{-5/2} \right] +  \delta\phi(r) \ ,
\end{align}
\ese
with
\be
\delta  \varphi_{i}(r) = c_{i}^{(+)} r^{\Delta_+} + c_{i}^{(-)} r^{\Delta_-} + k_\mt{UV}^2 \beta_i \, r^{\Delta_0} 
\ee
and we have collectively denoted 
\be
\{\delta \varphi\} \equiv  \{\delta  g_{tt}, \delta  g_{xx}, \delta  g_{{\mathbb {CP}}^3} , \delta \phi \} \ .
\ee
The set of constants $\left\{c_{i}^{(+)}, c_{i}^{(-)}\right\}$ is determined in terms of two free parameters, the set of constants $\{\beta_i\}$ is completely determined by the equations of motion,  and the exponents are given by
\bse
\bal
 \Delta_+^1 & = - 3 + \sqrt{\frac{3}{11} \left(381-4 \sqrt{793}\right)} \ , \\
 \Delta_+^2 & = - 3 + \sqrt{\frac{3}{11} \left(381+4 \sqrt{793}\right)} \ , \\
  \quad \Delta_0 & = 4 \ .
\end{align}
\ese
The modes $\Delta_{+}^{1,2}$ are exactly the ones in \eqref{eq.D2-IRmodes}, when the $k_\mt{UV}$-flux is absent, whereas the $\Delta_0$ mode encodes the presence of this flux. The fact that the latter decays as $r\to 0$ shows that indeed the $k_\mt{UV}$-flux induces a subleading correction to the IR solution \eqq{eqnspl}. 

These results suggest the existence of a full RG flow from the $k_\mt{UV}$-dominated solution \eqq{abjmUV} in the UV to the quark-density-dominated solution  \eqq{eqnspl} in the IR. We leave the explicit construction of this flow for future work.

\section{Discussion}\label{sec.discussion}

In this paper we have constructed the gravity duals of quantum field theories in various dimensions in the presence of a non-zero density of external, heavy quarks.  The new scale introduced by the quark density induces an RG  flow to a new IR characterised by hyperscaling violating and dynamical exponents, where the scaling symmetry is mildly broken by  logarithmically-running scalars, as shown in section \ref{IRsolutions}.
In particular, the charge density determines a crossover scale, $\uc$, at which a transition  between the UV and IR regimes  occurs, given by
\be
\uc \sim \lambda^\frac{2}{6-p}  \, \nq^\frac{1}{6-p} \ .
\ee
The relation among $\uc$ and the  scales present in the uncharged quantum field theory gives rise to different classes of RG flows, discussed at length in Section \ref{scales}.

In the main text we have focused on holographic duals of large-$\nc$, $d$-dimensional SYM theories (for $d\leq 6$), dual to the near-horizon geometries of stacks of D$p$-branes  at the tip of  $(9-p)$-dimensional cones in type II supergravity. 
When the base of the cone is an $(8-p)$-sphere the gauge group of the SYM theory is $U(N)$.
Our results do not depend on the explicit details of the compact $(8-p)$-manifold as long as this is an Einstein manifold. Therefore our construction includes the cases of theories with different gauge groups and matter in the adjoint and bifundamental representations. For example, in the case $p=3$,  gauge theories dual to geometries with cones on a base with topology $S^2\times S^3$ are described in  \cite{Martelli:2004wu,Benvenuti:2004dy,Benvenuti:2005ja,Franco:2005sm,Butti:2005sw}.

We emphasize that the compact manifold needs not admit Killing spinors, so supersymmetry is not a requirement of our construction. Nevertheless, it includes many supersymmetric cases beyond the simple SYM theories that we have focused on. One example is provided by the  $\tfrac{1}{4}$-BPS F1-D3 intersection in  \cite{Faedo:2013aoa}, where the IR Lifshitz solution with $z=7$ of \cite{Kumar:2012ui} was also found. 

Although supersymmetry is not a determining factor in the existence of the IR solutions \eqref{IRgeneric}--\eqref{eq.genericscalingend}, it ensures the stability of the starting point, namely of the uncharged solution. This however does not guarantee the stability of the IR solutions, since the addition of charge breaks supersymmetry. We have shown that there are no obvious thermodynamic instabilities associated for example to a negative specific heat, but it would be interesting to study the spectrum of fluctuations around the solution and check for the existence of tachyonic modes.

One  physical quantity that would be interesting to study in future work is the entanglement entropy. In holography, this is obtained from the area of an extremal surface extending into the bulk and anchored at the boundary of the spacetime at the entangling region \cite{Ryu:2006bv}.

Theories with a hyperscaling-violating exponent satisfying the relation $\theta=p-1$ were shown in  \cite{Ogawa:2011bz} to give rise to a logarithmic violation of the entanglement entropy. 
This was originally interpreted as a smoking gun of hidden fractional Fermi surfaces  \cite{Huijse:2011ef}. However, the results of \cite{Hartnoll:2012wm} showed that despite this logarithmic violation of the entanglement entropy the transverse current spectral density\footnote{To obtain this quantity one needs to add a Maxwell term to the action.} displays an exponential suppression at finite momentum, even when the finite momentum excitations of the Fermi surface are scaled towards $\omega=0$.
This means that there are no low-energy degrees of freedom at finite momentum, and the system is bosonic.
In the present paper this situation is realised for $p=2$, which in the IR is described by a HV-Lif metric with $z=5$ and $\theta=1$.
If the typical size of the entangling region is large, and the string density is also sufficiently large, this extremal surface will be sensitive to the IR HV-Lif metric, and we expect to observe such a logarithmic violation.

There is a  case in which the transverse current spectral density vanishes at low energies as a momentum-dependent power and, at the same time, the entanglement entropy is logarithmically violated.
This is  an indication that there are low-energy degrees of freedom at finite momentum, and suggests a fermionic behavior of the dual theory.
To obtain this behavior of the spectral density one must consider the $z\to\infty$ limit with $z/\theta=-1$ \cite{Hartnoll:2012wm}.
This is precisely the limit in which the IR fixed point of 4+1 SYM theory with external quark density is obtained.
It would be interesting to relate these facts with the  $AdS_3\times\mathbb{R}^4$ geometry emerging in the IR, in a similar way as the flow from $AdS_5$ to $AdS_3\times\mathbb{R}^2$ in \cite{D'Hoker:2009mm} was due to the fermionic zero modes governing the end point of the RG flow.

Since the construction in this paper is stringy in nature, it is possible to probe the geometries we have constructed with D$p$-branes in the quenched approximation, where the backreaction on the geometry of the newly added brane is neglected.
The results in \cite{Hartnoll:2009ns,HoyosBadajoz:2010kd} are not of direct use here, since in our top-down construction there is a non-trivial dilaton and background Ramond-Ramond  forms that can give rise to non trivial Wess-Zumino terms in the D$p$-brane action. The simplest setup would be to consider the probe brane in the IR HV-Lif geometry of section \ref{IRsolutions}. A study of probe branes along the domain wall solutions of section \ref{sec.flows} is also possible, but this requires working with a  numerical background.

\section*{Acknowledgements}

We thank Blaise Gouteraux, Kostas Skenderis and specially Prem Kumar for discussions. 
We are supported by grants 2014-SGR-1474, MEC FPA2010-20807-C02-01, MEC FPA2010-20807-C02-02, CPAN CSD2007-00042 Consolider- Ingenio 2010, and ERC Starting Grant ÒHoloLHC-306605Ó. JT is also supported by  the Juan de la Cierva program of the Spanish Ministry of Economy.

\appendix

\section{Ten-dimensional equations and dimensional reduction}
 To perform the dimensional reduction we find convenient to start with the 10D action in Einstein frame, which reads
\bal
S & = \frac{1}{2\kappa_{10}^2} \int \Big[ {\cal R}\, *1-\frac{1}{2} \d \phi \wedge *\d \phi - \frac{1}{2} e^{-\phi} H \wedge *H \ -\frac{1}{2} e^{\frac{5-p}{2}\phi} F_p \wedge * F_p, \nonumber \\
& \qquad\qquad -\frac{1}{2} e^{\frac{p-3}{2}\phi} F_{8-p} \wedge * F_{8-p} \Big] - \frac{N_q}{2\pi\ell_s^2} \int \left( e^{\frac{\phi}{2}}{\cal K} - B \right) \wedge \omega_8 \ . \label{eq.10Daction}
\end{align}
The last factor is the Nambu-Goto action for the $N_q$ strings, with
\bal
{\cal K} & = \sqrt{-G_{tt} \, G_{rr}} \, \d t \wedge \d r = e^{\frac{2(8-p)}{p}\eta} \, \sqrt{-g_{tt} \, g_{rr}} \, \d t \wedge \d r \ , \\
\omega_8 & = \frac{1}{V_{8-p}} \d x^1 \wedge \cdots \wedge \d x^p \wedge \omega_{8-p} \ .
\end{align}
In these equations we are implicitly assuming that the 10D metric takes the form
\be
\d s_{10}^2 (G) = e^{-\frac{2(8-p)}{p}\eta} \d s_{p+2}^2 (g) + e^{2\eta} L^2 \d\Omega_{8-p}^2 \ ,
\ee
with
\be
\d s_{p+2}^2 (g)= g_{tt} \, \d t^2 + g_{xx} \, \d x_p^2 + g_{rr}\, \d r^2 
\ee
and $\d \Omega_{8-p}^2$ the line element of an Einstein manifold of unit radius. In particular, its Ricci scalar is ${\cal R}[\Omega_{8-p}]=(8-p)(7-p)/L^2$ and  its volume is $\int \omega_{8-p} = V_{8-p}$.

Before discussing the dimensional reduction, let us present the $10$-dimensional equations of motion (in Einstein frame) that we are solving. The equation of motion resulting from the variation of the NSNS two-form $B$ will play a crucial role, thus we first pause to offer a few comments. Let us recall that the complete action for Type IIA/IIB supergravity, when  $H=dB$  is included, contains more terms than equation (\ref{eq.10Daction}). First of all, there is a Chern-Simons term containing $B$. In Type IIA this takes the form $- \frac{1}{4 \kappa_{10}^2} B\wedge F_4 \wedge F_4$ and in Type IIB it takes the form $- \frac{1}{4 \kappa_{10}^2} C_4 \wedge H \wedge F_3$. Second, the RR-field strengths are also redefined, subsequently denoted by $\tilde{F}_p$, including linear contributions from the two-form $B$. Thus the variation with respect to $B$ will receive contributions from the kinetic term for $H$, the Chern-Simons term and finally the kinetic term of the redefined RR-field strengths.

Furthermore, the Bianchi identities satisfied by the corresponding $F_p$ can also receive contribution from $H$. However, in this paper we will consider $H=0$ and therefore we provide the equations of motion below without any explicit contribution coming from $H$, in particular we present them in terms of $F_p$ instead of $\tilde{F}_p$. We make only one exception when we write down the equation of motion \eqq{pepe1} or \eqq{pepe2} coming from the variation of the action with respect to $B$, where we chose to include the $H$-contribution coming from the kinetic term for $B$ to remind the reader about the origin of this equation.

Now, the Bianchi identities are:
\begin{eqnarray}
\d F_p =  0 \ , \quad \d F_{8 - p} = 0 \ .
\end{eqnarray}
The equation of motion for the NSNS $2$-form $B$ is best written separately for Type IIA and Type IIB as:
\begin{eqnarray}
\d \left( e^{-\phi} \ast H \right)  - e^{\frac{\phi}{2}} F_2 \wedge \ast F_4 + \frac{1}{2} F_4 \wedge F_4 & = & - \frac{2 \kappa_{10}^2}{2\pi \ell_s^2} N_q \, \omega_8 \ , \quad \left[ {\rm type \, IIA} \right] \label{pepe1} \\
\d \left( e^{- \phi} \ast H \right)  + e^{\phi} F_1 \wedge \ast F_3 + \frac{1}{2} F_3 \wedge F_5 & = & - \frac{2 \kappa_{10}^2}{2\pi \ell_s^2} N_q \, \omega_8 \ , \quad \left[ {\rm type \, IIB} \right] \ . \label{pepe2}
\end{eqnarray}
As we have explained, the first, the second and the third term of the above equations respectively originate from the variation of the kinetic term of $H$, the kinetic term of the $\tilde{F}$ RR-fluxes and the Chern-Simons term of the corresponding supergravity.

The equations of motion for the RR-fluxes\footnote{For Type IIB supergravity, we need to further impose the self-duality condition: $F_5 = \ast F_5$.} and the dilaton are given by
\begin{eqnarray}
\d \left( e^{\frac{5-p}{2} \phi}\ast F_p \right) & = & 0  \ , \\
\d \left ( e^{\frac{p-3}{2} \phi} \ast F_{8-p} \right) & = & 0 \ , \\
 \d \ast \d \phi & = &  \frac{2\kappa_{10}^2 N_q}{2\pi \ell_s^2} \frac{1}{2} e^{\frac{\phi}{2}} {\cal K} \wedge \omega_8 + \frac{5-p}{4} \left( e^{\frac{5-p}{2} \phi} F_p \wedge \ast F_p \right) \nonumber\\
 & + & \frac{p-3}{4} \left( e^{\frac{p-3}{2} \phi} F_{8-p} \wedge \ast F_{8-p} \right) \ .
\end{eqnarray}
Finally the Einstein equations are:
\begin{eqnarray}
{\cal R}_{AB} - \frac{1}{2} G_{AB} {\cal R}  = T_{AB}^{\rm sugra}  + T_{AB}^{\rm strings} \ ,
\end{eqnarray}
with
\begin{eqnarray}
T_{AB}^{\rm sugra} & = & T_{AB}^{\left(\phi \right)} + T_{AB}^{\left(F_p\right)} + T_{AB}^{\left(F_{8-p}\right)} \ , \nonumber \\
T_{AB}^{\left(F_p\right)} & = & \frac{1}{2}  e^{\frac{5-p}{2}\phi} \left( \iota_{(A} \left(F_p \right) \lrcorner \iota_{B)} \left(F_p \right) - \frac{1}{2} G_{AB} \left( F_p \lrcorner F_p \right) \right) \ , \\ 
T_{AB}^{\left(F_{8-p}\right)} & = & \frac{1}{2} e^{\frac{p-3}{2}\phi} \left( \iota_{(A} \left(F_{8-p} \right) \lrcorner \iota_{B)} \left( F_{8-p} \right)  - \frac{1}{2} G_{AB} \left ( F_{8-p} \lrcorner F_{8-p} \right) \right) \ , \\
T_{AB}^{\left(\phi \right)} & = & \frac{1}{2} \left[ \left(\partial_A \phi \right) \left(\partial_B \phi \right) - \frac{1}{2} G_{AB} \left( \d \phi \lrcorner \d \phi \right) \right] \\
T_{AB}^{\rm strings} & = & \frac{2 \kappa_{10}^2 N_q}{2 \pi \ell_s^2} \, \frac{1}{2} e^{\frac{\phi}{2}}  \left[ G_{AB} \,  \omega_8 \lrcorner (\ast {\cal K}) - \iota_{(A} \omega_8 \lrcorner \iota_{B)} \left(* {\cal K} \right) \right] \ ,
\end{eqnarray}
where $G$ stands for the Einstein frame metric and we use, for any $p$-forms $\Gamma_p$ and $\Delta_p$, the following definitions
\begin{eqnarray}
\iota_A \Gamma_p & = & \frac{1}{(p-1)!}(\Gamma_p)_{A C_1\cdots C_{p-1}} \,\d x^{C_1} \wedge \cdots \wedge \d x^{C_{p-1}} \ , \nonumber \\
\Gamma_p \lrcorner \Delta_p & = & \frac{1}{p!}(\Gamma_p)_{A_1\cdots A_p} (\Delta_p)^{A_1\cdots A_p} \ .
\end{eqnarray}

Now let us discuss the dimensional reduction of the D$p$+strings system on the compact $8-p$  manifold. As explained in the main text, the presence of D$p$-branes is encoded in an $(8-p)$ RR form
\be \label{eq.F8mp}
F_{8-p} = (7-p) \, L^{7-p} \, \omega_{8-p} \ ,
\ee
and the  strings source a component
\be
F_p = (-1)^{\left[\frac{p+1}{2} \right]} \frac{{Q}}{L} \, \d x^1 \wedge \cdots \wedge \d x^p \ ,
\ee
with $Q$ a constant to satisfy the Bianchi identity (eventually we will set $H=0$ and the Bianchi is $\d F_p=0$). The Ricci scalar of the 10D metric can be expressed in terms of the $(p+2)$-dimensional Ricci scalar as
\be
{\cal R}[G]  = e^{\frac{2(8-p)}{p}\eta} \, {\cal R}[g] + \frac{e^{-2\eta}}{L^2} \, (8-p)(7-p) -  \frac{8(8-p)}{p} \, e^{\frac{2(8-p)}{p} \eta} \, \partial_\mu \eta \partial^\mu \eta  + \frac{2(8-p)}{p} \, e^{ \frac{2(8-p)}{p} \eta} \, \Box \eta \ .
\ee

With this ansatz and a radius-dependent dilaton we have all the necessary fields to describe the system.
The equation of motion for the NSNS field, with the ansatz $B=0$, relates the constant $Q$ to the number of strings as 
\be
\frac{N_q}{2\pi \ell_s^2}  = \frac{V_{8-p}}{2\kappa_{10}^2} (7-p)\, L^{6-p} {Q} \ ,
\ee
where  the equations of motion of the RR forms have been used. 

To integrate  on the compact manifold we need the following information
\bal
*1 & = L^{8-p} \,e^{-\frac{2(8-p)}{p}\eta} \, (* 1)\wedge \omega_{8-p} \ , \\
F_q \wedge * F_q & = L^{8-p} \,e^{-\frac{2(8-p)}{p}\eta} \, F_q \lrcorner F_q \, (* 1)\wedge \omega_{8-p} \ ,
\end{align}
where the $*$ operator is the 10D one in the left-hand side of the equations and the $(p+2)$-dimensional one in the right-hand side. Notice, however, that the contraction $\lrcorner$ is the 10D one. In particular
\bal
F_p \lrcorner F_p & = \frac{Q^2}{L^2} \, e^{2(8-p)\eta} \, g_{xx}^{-p} \ , \\
F_{8-p} \lrcorner F_{8-p} & = (7-p)^2 \, L^{6-p} \, e^{-2(8-p)\eta} \ .
\end{align}

Expanding the action  and integrating over the compact manifold we obtain, up to total derivatives,
\be
S  = \frac{1}{2\kappa_{p+2}^2}  \int \d x^{p+2} \sqrt{-g} \left[ {\cal R}[g] - \frac{8(8-p)}{p} \, \partial_\mu \eta \partial^\mu \eta \, - \frac{1}{2} \partial_\mu \phi \partial^\mu \phi - V(\eta,\phi) \right] \ ,
\ee
with potential
\be
V  = \frac{1}{2\,L^2} \left( \frac{Q}{g_{xx}^{p/2}} \, e^{\frac{5-p}{4}\phi}\, e^{\frac{(8-p)(p-1)}{p}\eta}  +  (7-p) \, e^{\frac{p-3}{4}\phi} \, e^{\frac{(p-8)(p+1)}{p} \eta} \right)^2 - \frac{(8-p) \, (7-p)}{L^2} e^{-\frac{16}{p} \eta}  \ .
\ee
The kinetic terms involving $F_p$ and $F_{8-p}$ have completed a square with the Nambu-Goto action, which corresponds to the crossed term in this square.
This action has appeared before for $p=3$ (and setting $L=1$) in \cite{Kumar:2012ui}.
We have defined the reduced Newton's constant as
\be
\frac{1}{2\kappa_{p+2}^2} = \frac{L^{8-p} \, V_{8-p}}{2\kappa_{10}^2} \ .
\ee

The solutions in section \ref{sec.solutions} follow from the equations of motion derived from \eqref{eq.pplus2daction}.
For the D$p$-branes' solution in the absence of strings we can truncate the system with the identification
\be\label{eq.etaphiidentification}
\eta = \frac{p-3}{4(7-p)} \phi \  ,
\ee
and the action takes the form of the one given in \cite{Boonstra:1998mp}.

\vspace{10pt}

For the IR solutions we can obtain the scaling coefficients for generic $p$ straightforwardly.
The Einstein equations of motion can be manipulated such that we can express ${\cal R}[g]\sim V$.
Therefore, the hyperscaling coefficient, together with the radial dependence of the scalars, can be deduced assuming that all the terms in the potential plus the Ricci scalar go with the same power of the radial coordinate.
Given that the Ricci scalar for the Lifshitz and hyperscaling violating ansatz behaves as ${\cal R}\sim r^{\frac{2\theta}{p}}$, we obtain the values quoted in \eqref{eq.IRscalars} and \eqref{eq.genericscalingend}.
On the other hand, the value of the dynamical exponent $z$ is obtained by considering the combination of the $tt$ and $rr$ components of Einstein's equation that does not contain the potential, which gives an algebraic equation for $z$ in terms of $p$ and $\theta$.
This procedure gives  the right values of $z$ and $\theta$ provided a solution exists. Notice however that there are other algebraic equations to be satisfied, coming from the rest of the equations of motion.

\end{document}